\documentclass[aps,pra,showpacs,twocolumn,nofootinbib]{revtex4-2}
\usepackage{bm}

\usepackage{longtable}

\usepackage{amsmath}
\usepackage{amssymb}
\usepackage{slashed}
\usepackage{float}
\allowdisplaybreaks
\usepackage{dcolumn}
\usepackage{graphicx}

\usepackage{nicefrac}
\newcolumntype{w}[1]{D{.}{.}{#1}}
\newcommand{\Za}{Z\alpha}
\newcommand{\vare}{\varepsilon}
\newcommand{\balpha}{\bm{\alpha}}

\newcommand{\lbr}{\langle}
\newcommand{\rbr}{\rangle}

\newcommand{\bra}[1]{\langle #1 |}
\newcommand{\ket}[1]{| #1 \rangle}

\newcommand{\intinf}{\int^{\infty}_{-\infty}}

\newcommand{\bfr}{{\bm r}}
\newcommand{\bfp}{{\bm p}}

\newcommand{\bfx}{{\bm x}}

\newcommand{\bfz}{{\bm z}}
\newcommand{\bfy}{{\bm{y}}}

\newcommand{\I}[4]{I_{#1 #2\, #3 #4}}

\begin{document}

\title{QED calculations of the $\bm{2p}\,$-$\,\bm{2s}$ transition energies in Li-like ions}

\author{V.~A. Yerokhin}
\affiliation{Max~Planck~Institute for Nuclear Physics, Saupfercheckweg~1, D~69117 Heidelberg, Germany}

\author{Z. Harman}
\affiliation{Max~Planck~Institute for Nuclear Physics, Saupfercheckweg~1, D~69117 Heidelberg, Germany}

\author{C.~H. Keitel}
\affiliation{Max~Planck~Institute for Nuclear Physics, Saupfercheckweg~1, D~69117 Heidelberg, Germany}

\begin{abstract}

Systematic QED calculations of ionization energies of the
$2s$, $2p_{1/2}$, and $2p_{3/2}$ states, as well as the $2p_{1/2}$--$2s$ and $2p_{3/2}$--$2p_{1/2}$
transition energies
are performed for Li-like ions with the nuclear charge numbers $Z = 10$--$100$.
The convergence of QED perturbative expansion is improved by using the
extended Furry picture, which starts from the Dirac equation with a local screening potential.
An {\em ab initio} treatment is accomplished for
one- and two-photon electron-structure QED effects
and the one-photon screening of the self-energy and vacuum-polarization corrections.
This is complemented with an approximate treatment of the two-photon QED screening
and higher-order (three or more photon) electron-structure effects.
As a result,
the obtained theoretical predictions improve upon the accuracy achieved in previous
calculations.
Comparison with available experimental data shows a good agreement between
theory and experiment.  In most cases, the theoretical values surpass the experimental
results in precision, with only a few exceptions.
In the case of uranium and bismuth,
the comparison provides one of the most stringent tests of bound-state QED in
the strong-field regime. Alternatively, the obtained
results can be employed for high-precision
determinations of nuclear charge radii.

\end{abstract}

\maketitle

\section*{Introduction}

Lithium-like ions are among the simplest atomic systems, and their spectra can be described with high accuracy
using modern {\em ab initio} theoretical methods.
Although they contain more electrons than H- and He-like ions --- and are therefore more challenging for
theoretical description
 --- Li-like ions turn out to be more accessible for experimental studies.
For example, in heavy one- and two-electron ions, the $K$-shell transition energies lie in the hard X-ray range, making the
detection of emitted radiation technically very demanding \cite{gumberidze:05:prl,loetzsch:24,pfafflein:25}.
In contrast, the $2p\,$-$\,2s$ transition lines in heavy Li-like ions fall within the softer X-ray region,
where significantly higher experimental accuracy has been achieved \cite{beiersdorfer:93,beiersdorfer:98,beiersdorfer:05}.
Li-like ions
therefore represent an attractive compromise between the feasibility of high-precision {\em ab initio} theoretical
treatment and the practicality of accurate experimental measurement.

Further experimental advances in the spectroscopy of Li-like ions are anticipated in the near future.
In light Li-like ions, the $2p\,$-$\,2s$ transitions lie in the extreme ultraviolet (XUV) region.
The project of developing the XUV frequency comb \cite{nauta:17,nauta:21} aims to enable
spectroscopy of such transitions with unprecedented accuracy.
In the high-$Z$ regime, a measurement of the $2p_{1/2}$--$2s$ transition
in lithium-like lead is planned as a proof-of-principle experiment for precision X-ray
spectroscopy within the Gamma Factory project at CERN~\cite{budker:20}.

Different theoretical methods are employed for {\em ab initio} calculations of Li-like ions,
depending on the nuclear charge $Z$.
For light atoms, the most powerful current approach
is based on nonrelativistic quantum electrodynamics (NRQED), which expands
energies in powers of $\alpha$ and $Z\alpha$, where $\alpha$ is the fine-structure
constant. Highly advanced NRQED calculations were performed by
Puchalski and Pachucki for the lowest-lying states of Li and Be$^+$
\cite{puchalski:08:li,puchalski:14,puchalski:15}.

For heavy ions, the best results are
obtained within the QED approach that treats the nuclear binding strength parameter
$Z\alpha$ to all orders, while expanding in the electron-electron interaction, characterized
by the parameter $1/Z$.
Calculations based on this method have been performed for Li-like ions by the St.~Petersburg group \cite{yerokhin:01:2ph,yerokhin:06:prl,yerokhin:07:lilike,kozhedub:10} and
the Notre Dame group \cite{sapirstein:01,sapirstein:11}.
Comparison of results of these calculations with high-precision measurements of Li-like bismuth and uranium \cite{beiersdorfer:93,beiersdorfer:98,beiersdorfer:05}
has provided some of the most stringent tests of bound-state QED in the nonperturbative regime with respect to
the nuclear binding strength.

Since the theoretical calculations of Li-like ions reported in
Refs.~\cite{kozhedub:10,sapirstein:11}, advanced
theoretical techniques have been developed for a more accurate treatment of electron-correlation
and QED screening effects \cite{kosheleva:22,malyshev:23,malyshev:24:belike,morgner:25}. The aim of the
present work is to apply these advancements to the calculation of energy levels of Li-like ions,
thereby improving the accuracy of theoretical predictions.

The paper is organized as follows. Sec.~\ref{sec:struct} outlines the theoretical treatment
of the electron-structure effects, i.e., those arising solely from the electron-electron interaction
and excluding radiative corrections with closed loops. First, Sec.~\ref{sec:DCB}
addresses  the solution of the Dirac-Coulomb-Breit Hamiltonian within the no-pair approximation.
This is complemented in Sec.~\ref{sec:structqed}
by a separate evaluation of
QED electron-structure effects
induced by the exchange of one and two virtual photons
between the electrons.  Sec.~\ref{sec:structnum} summarizes numerical results for
the electron-structure part of the energies.
Sec.~\ref{sec:qed} describes the theoretical treatment of radiative QED effects. Its first part,
Sec.~\ref{sec:qed1el}, focuses on the one-electron QED contributions. Then Sec.~\ref{sec:qedscr} details
the evaluation of QED screening corrections. The QED treatment of the nuclear recoil effect
is presented in Sec.~\ref{sec:rec}, while the nuclear effects are discussed in Sec.~\ref{sec:nucl}.
Sec.~\ref{sec:results} summarizes the total theoretical results for the ionization and
transition energies of Li-like ions with $Z = 10$--$100$ and compares them with available experimental
results. Finally, Sec.~\ref{sec:qedmod} compares {\em ab initio} QED calculations with
the approximate treatment based on the model QED operator.

Relativistic units $\hbar = c = 1$ and charge units $\alpha = e^2/(4\pi)$ are used throughout
this paper.

\section{Electronic structure}
\label{sec:struct}

\subsection{Dirac-Coulomb-Breit energy}
\label{sec:DCB}

The Dirac-Coulomb-Breit (DCB) Hamiltonian of an $N$-electron atom can be written as
\begin{align}\label{eq1}
    H_{\rm DCB} = &\ \sum_i \Big[
    \balpha_i\cdot\bfp_i+ (\beta-1)\,m+ V_{\rm nuc}(r_i)
    \Big]
\nonumber \\    &
    + \sum_{i<j} \Lambda_{++}\, I(0,r_{ij})\,\Lambda_{++}\,,
\end{align}
where indices $i,j = 1,\ldots,N$ numerate the electrons,
$\balpha$ and $\beta$ are the Dirac matrices, $V_{\rm nuc}$ is the electrostatic
potential of the nucleus,
$I(0,r_{ij})$ is the electron-electron interaction operator in the Breit approximation,
\begin{equation}
    I(0,r_{ij}) =  \frac{\alpha}{r_{ij}} -\frac{\alpha}{2r_{ij}}
   \Big[ \balpha_i\cdot\balpha_j + (\balpha_i\cdot\hat{\bfr}_{ij})(\balpha_j\cdot\hat{\bfr}_{ij}) \Big]\,,
   \label{eq:3}
\end{equation}
and $\hat{{\bm r}} = {\bm r}/|\bm{r}|$.
It should be noted that $I(0,r_{ij})$ is the zero-frequency limit of the Coulomb-gauge
electron-electron interaction operator in full QED, which will be introduced in
the next section.

We assume that the Hamiltonian $H_{\rm DCB}$ acts in the space spanned by
products of eigenstates of one-electron Dirac-Coulomb Hamiltonian
$h_D$ 
\begin{equation}\label{eq4}
    h_D(r) = \balpha\cdot\bfp+ (\beta-1)\,m+ V_{\rm nuc}(r) + U(r)\,,
\end{equation}
where $U(r)$ is a screening potential that approximately accounts for the effects of the other electrons.
Furthermore, $\Lambda_{++}$ is the 
projection operator onto the space spanned by products of the {\em positive-energy} eigenfunctions of $h_D$.  

It is important to note that the definition of the DCB Hamiltonian is not unique as it depends
on the choice of the screening potential $U(r)$ in $h_D$, which
defines the projector $\Lambda_{++}$.
Different definitions of the potential $U(r)$
will lead to DCB energies that vary due to differences in the omitted negative-energy contributions,
which induce corrections of the same order as QED effects,
see Ref.~\cite{sapirstein:99} for detailed discussion.

In the present work we use three types of local screening potentials
$U(r)$.
The simplest choice is the
core-Hartree (CH) potential induced by the charge density of core electrons,
\begin{equation}
    U_{\rm CH}(r) = \alpha \int_0^{\infty} dr^{\prime} \frac1{r_>}\, \rho_c^{s.c.}(r^{\prime})\,,
\end{equation}
where
$r_> = \max(r,r^{\prime})$, $\rho_c$ is the charge density of the core
electrons, and
the superscript ``$s.c.$" indicates that the density has to be calculated self-consistently.
The second type of potentials \cite{sapirstein:02} is taken from the density-functional theory (DFT)
\cite{kohn:65,cowan},
\begin{eqnarray}\label{eq6}
    U_{\rm DFT}(r) &=& \alpha \int_0^{\infty} dr^{\prime} \frac1{r_>}\,
        \rho_t^{s.c.}(r^{\prime})
 \nonumber \\ &&
     {} - x_{\alpha}\, \frac{\alpha}{r}\,
          \left[ \frac{81}{32\, \pi^2}\,r\,\rho_t^{s.c.}(r) \right]^{1/3}  \,,
\end{eqnarray}
where $\rho_t(r) = \rho_c(r)+\rho_v(r)$ is the total (core plus valence) electron
charge density, and $x_{\alpha} \in [0,1]$ is a free parameter.
The DFT potential with $x_{\alpha} = 2/3$ is called the Kohn-Sham (KS) potential,
whereas the one with $x_{\alpha} = 1$ is referred to as the Dirac-Slater (DS)
potential.
We note that in this work we do {\em not} apply the so-called Latter correction
of the asymptotics in the DFT potentials \cite{latter:55}. The reason is
that this correction
spoils the numerical stability of
our calculations of the self-energy screening corrections.

The third type of the screening potential used in this work is the local Dirac-Fock (LDF)
potential, obtained by inverting solutions of the Dirac-Fock equation for the valence
state $v$
\cite{shabaev:05:pra,yerokhin:08:pra}.
Since our primary interest are transition energies between different valence states,
it is advantageous to use {\em the same} LDF potential for all of them, because in
this case the contribution induced by the core electrons cancels identically.
So, in this work we use the LDF potential generated for the $2p_{1/2}$ valence state
for computations for all valence states. The additional advantage is that $2p_{1/2}$
wave function does not have any nodes, so that LDF potential does not need any smoothing.

Within the many-body perturbation theory (MBPT), the DCB Hamiltonian is represented
as a sum of the non-perturbed Hamiltonian $H_0$ and the interaction $H_I$,
$H_{\rm DCB} = H_0 + H_I$, where
\begin{align}
    H_0 =  \sum_i h_D(r_i) &\,\,,  \\
    H_I =
\sum_{i<j} \Lambda_{++}\, I(0,r_{ij})\,\Lambda_{++} &\,
 - \sum_{i} \Lambda_{+} \, U(r_i)\,\Lambda_{+}
    \,.
\end{align}
The DCB energy is obtained by applying the Rayleigh-Schr\"odinger perturbation theory
with $H_I$ as a perturbation. This leads to a perturbation expansion for the
DCB energy
\begin{equation}
E_{\rm DCB} = E^{(0)}+ E^{(1)} + E^{(2)} + E^{(3)}+\ldots\,.
\end{equation}
For the electronic configuration of one valence electron over one or several closed shells,
formulas for the perturbation corrections $E^{(i)}$ with $i \le 3$
were obtained in Ref.~\cite{blundell:87:adndt}. For the {\em ionization} energy of the
valence state $v$, one obtains \cite{blundell:87:adndt}
\begin{subequations}
\begin{align} \label{eq:20}
E^{(0)} =& \ \vare_v\ ,\\
E^{(1)} =& \big(V_{\rm HF}-U\big)_{vv}\,,  \label{eq:20:b} \\
E^{(2)} =&\  \sum_{amn} \frac{I_{vamn}\, I_{mn;va}}{\epsilon_{av}-\epsilon_{mn}}
 - \sum_{abm} \frac{I_{abmv}\, I_{mv;ab}}{\epsilon_{ab}-\epsilon_{vm}}
 \nonumber \\
 +& 2\,\sum_{am} \frac{(V_{\rm HF}-U)_{am}\, I_{mv;av}}{\vare_a-\vare_m}
 \nonumber \\
 +& \sum_{i\neq v} \frac{(V_{\rm HF}-U)_{vi}\, (V_{\rm HF}-U)_{iv}}{\vare_v-\vare_i}\,.
 \label{eq:21}
\end{align}
\end{subequations}
The standard MBPT conventions are used here:
the letter $v$ stands for the valence orbital;
the letters $a$, $b$, $c$, $\ldots$ designate occupied core orbitals;
$n$, $m$, $r$, $\ldots$ signify excited orbitals outside the core including the
valence orbital; $i$, $j$, $k$, $\ldots$ can be either excited or occupied orbitals.
All orbitals are positive-energy Dirac states.
Furthermore, $\vare_i$ is the Dirac energy of the state $i$,
$\epsilon_{ab} \equiv \vare_a + \vare_b$, the matrix elements are defined by
$I_{ab;cd} \equiv I_{abcd} - I_{abdc}$, $I_{abcd} \equiv \lbr ab|I(0)|cd\rbr$,
$U_{ab} = \lbr a|U|b\rbr$,
and $I(0)$ is the operator of the electron-electron interaction defined in Eq.~\eqref{eq:3}.
Moreover, the matrix elements $( V_{\rm HF} )_{ij}$ are defined by
\begin{align}
( V_{\rm HF} )_{ij} = \sum_a I_{ai;aj} \,.
\end{align}

The expressions for the third-order MBPT correction
$E^{(3)}$ are rather lengthy; they can be found in Refs.~\cite{blundell:87:adndt,johnson:88:b}
and will not be repeated here. Alternative formulas for the third-order MBPT correction were derived
in Ref.~\cite{yerokhin:07:lilike}. In the present work we adopt formulas from
Ref.~\cite{blundell:87:adndt} since they turned out to be more stable numerically than those
from Ref.~\cite{yerokhin:07:lilike}.

It might be noted that in the particular case of the screening potential $U$ being
the Dirac-Fock potential,
formulas for the MBPT corrections simplify greatly. For example,
the correction $E^{(1)}$ and
the last two terms
in Eq.~(\ref{eq:21}) vanish. However, we do not use the Dirac-Fock potential here,
because the Furry picture of QED can be formulated only for local potentials.
In order to keep the DCB energy compatible to the QED part of
our calculations, we use only local potentials $U$ and employ the full
expressions for the MBPT corrections.

The second method used in this work for computing the DCB energies
is the configuration-interaction (CI) method.
In this method,
the $N$-electron wave function of the atom with parity $P$, angular momentum
quantum number $J$, and momentum projection $M$ is represented as a linear combination
of configuration-state functions (CSFs),
\begin{equation}
  \Psi(PJM) = \sum_r c_r \Phi(\gamma_r PJM)\,,
\end{equation}
where $\gamma_r$ denotes the set of additional quantum numbers that determine the
CSF. The CSFs are constructed as $jj$-coupled
antisymmetrized products of one-electron orbitals
$\psi_i$ which are positive-energy eigenfunctions of the one-particle Hamiltonian $h_D$.
In this way, we ensure that the $\Lambda$ projection operator in the CI method
is identical to that used in the MBPT calculations.

The DCB energies and the mixing coefficients
$c_r$ are obtained by solving the secular equation
\begin{equation}
    {\rm det} \bigl\{\lbr \gamma_r PJM|H_{\rm DCB}|\gamma_s PJM\rbr -E_r\,\delta_{rs}\bigr\} =
    0\,
\end{equation}
and determining the eigenvalues of
the Hamiltonian matrix.
Our implementation of the CI method
uses the one-electron basis constructed with $B$-splines
\cite{shabaev:04:DKB}. A
description of the numerical procedure can be found
in Refs.~\cite{yerokhin:08:pra,yerokhin:12:lilike}.

%
%
%
\begin{table*}[htb]
\begin{center}
\caption{
Dirac-Coulomb-Breit transition
energies (in a.u) calculated within the MBPT approach
and the CI method, with the LDF starting potential.
\label{tab:mbptci}
}
\begin{ruledtabular}
\begin{tabular}{llw{2.9}w{2.9}w{2.9}w{2.9}w{2.9}}
           & \multicolumn{3}{c}{$2p_{1/2}$--$2s$}
                &  \multicolumn{3}{c}{$2p_{3/2}$--$2s$}
\\
\cline{2-4}
\cline{5-7}
\\[-9pt]
$Z$   & \multicolumn{1}{c}{MBPT}
      & \multicolumn{1}{c}{CI}
      & \multicolumn{1}{c}{Diff.}
      & \multicolumn{1}{c}{MBPT}
      & \multicolumn{1}{c}{CI}
      & \multicolumn{1}{c}{Diff.}
\\\hline\\[-9pt]
 10    &  0.584\,588\,(44)   &  0.584\,572\,(6)    & -0.000\,016\,(44)   &  0.592\,093\,(41)   &  0.592\,076\,(6)    & -0.000\,016\,(41) \\
 11    &  0.657\,340\,(37)   &  0.657\,328\,(6)    & -0.000\,012\,(38)   &  0.669\,280\,(34)   &  0.669\,267\,(6)    & -0.000\,012\,(35) \\
 12    &  0.730\,327\,(31)   &  0.730\,318\,(6)    & -0.000\,009\,(31)   &  0.748\,421\,(28)   &  0.748\,411\,(6)    & -0.000\,009\,(28) \\
 13    &  0.803\,610\,(27)   &  0.803\,603\,(7)    & -0.000\,007\,(28)   &  0.829\,974\,(24)   &  0.829\,966\,(6)    & -0.000\,007\,(25) \\
 14    &  0.877\,244\,(23)   &  0.877\,238\,(7)    & -0.000\,006\,(24)   &  0.914\,431\,(20)   &  0.914\,425\,(7)    & -0.000\,006\,(21) \\
 15    &  0.951\,277\,(21)   &  0.951\,273\,(7)    & -0.000\,005\,(22)   &  1.002\,324\,(18)   &  1.002\,319\,(7)    & -0.000\,005\,(19) \\
 16    &  1.025\,758\,(18)   &  1.025\,754\,(4)    & -0.000\,004\,(19)   &  1.094\,223\,(15)   &  1.094\,219\,(7)    & -0.000\,005\,(17) \\
 17    &  1.100\,729\,(17)   &  1.100\,725\,(8)    & -0.000\,004\,(19)   &  1.190\,740\,(13)   &  1.190\,736\,(8)    & -0.000\,005\,(16) \\
 18    &  1.176\,236\,(15)   &  1.176\,232\,(9)    & -0.000\,004\,(17)   &  1.292\,532\,(12)   &  1.292\,527\,(8)    & -0.000\,005\,(14) \\
 19    &  1.252\,322\,(14)   &  1.252\,318\,(10)   & -0.000\,004\,(17)   &  1.400\,297\,(11)   &  1.400\,293\,(9)    & -0.000\,005\,(14) \\
 20    &  1.329\,027\,(13)   &  1.329\,023\,(10)   & -0.000\,004\,(16)   &  1.514\,781\,(9)    &  1.514\,776\,(9)    & -0.000\,005\,(13) \\
 21    &  1.406\,392\,(12)   &  1.406\,387\,(12)   & -0.000\,005\,(17)   &  1.636\,773\,(9)    &  1.636\,767\,(11)   & -0.000\,006\,(14) \\
 22    &  1.484\,462\,(12)   &  1.484\,457\,(13)   & -0.000\,005\,(17)   &  1.767\,112\,(8)    &  1.767\,106\,(11)   & -0.000\,006\,(14) \\
 30    &  2.139\,529\,(10)   &  2.139\,522\,(19)   & -0.000\,007\,(22)   &  3.227\,297\,(4)    &  3.227\,289\,(16)   & -0.000\,009\,(16) \\
\end{tabular}
\end{ruledtabular}
\end{center}
\end{table*}

%
%
%
\begin{table*}
\begin{center}
\caption{
Electron-structure MBPT ($E^{(0)}$, $E^{(1)}$, $E^{(2)}$, $E^{(3)}$)
and QED ($E^{(1)}_{\rm qed}$, $E^{(2)}_{\rm qed}$) contributions
to the ionization energies of the $2s$, $2p_{1/2}$, and $2p_{3/2}$ states of Li-like zinc ($Z = 30$)
and bismuth ($Z = 83$),
for different starting potentials, in a.u.
\label{tab:elstruct:1}
}
\begin{ruledtabular}
\begin{tabular}{lllw{2.9}w{2.9}w{2.9}w{2.9}w{2.9}}
 $Z$ & State & Term
           & \multicolumn{1}{c}{Coul}
                &  \multicolumn{1}{c}{CH}
                    &  \multicolumn{1}{c}{KS}
                    &  \multicolumn{1}{c}{DS}
                    &  \multicolumn{1}{c}{LDF}
\\ \hline\\[-9pt]
 %
  30 & $2s$        & $E^{(0)}$           &  -114.228\,111 &  -101.664\,092 &  -100.002\,071 &  -101.357\,465 &  -102.071\,621 \\
     &             & $E^{(1)}$           &    12.230\,112 &    -0.615\,808 &    -2.260\,112 &    -0.893\,229 &    -0.198\,409 \\
     &             & $E^{(2)}$           &    -0.268\,227 &     0.013\,791 &    -0.004\,495 &    -0.016\,240 &     0.003\,525 \\
     &             & $E^{(3)}$           &    -0.000\,362 &    -0.000\,562 &     0.000\,037 &     0.000\,299 &    -0.000\,147 \\
     &             & $E^{(1)}_{\rm qed}$ &     0.000\,105 &     0.000\,085 &     0.000\,088 &     0.000\,090 &     0.000\,087 \\
     &             & $E^{(2)}_{\rm qed}$ &     0.000\,080 &     0.000\,096 &     0.000\,088 &     0.000\,085 &     0.000\,094 \\
     &             & Sum                 &  -102.266\,402 &  -102.266\,489 &  -102.266\,465 &  -102.266\,459 &  -102.266\,471 \\
     & $2p_{1/2}$  & $E^{(0)}$           &  -114.228\,638 &   -99.739\,390 &   -97.497\,371 &   -98.975\,026 &  -100.176\,769 \\
     &             & $E^{(1)}$           &    14.514\,158 &    -0.394\,157 &    -2.621\,125 &    -1.126\,696 &     0.050\,447 \\
     &             & $E^{(2)}$           &    -0.411\,197 &     0.006\,753 &    -0.008\,740 &    -0.026\,050 &    -0.000\,743 \\
     &             & $E^{(3)}$           &    -0.001\,369 &    -0.000\,332 &     0.000\,110 &     0.000\,650 &    -0.000\,058 \\
     &             & $E^{(1)}_{\rm qed}$ &    -0.000\,089 &    -0.000\,077 &    -0.000\,080 &    -0.000\,082 &    -0.000\,079 \\
     &             & $E^{(2)}_{\rm qed}$ &     0.000\,106 &     0.000\,025 &     0.000\,046 &     0.000\,061 &     0.000\,037 \\
     &             & Sum                 &  -100.127\,029 &  -100.127\,177 &  -100.127\,159 &  -100.127\,143 &  -100.127\,165 \\
     & $2p_{3/2}$  & $E^{(0)}$           &  -112.839\,015 &   -98.624\,169 &   -96.390\,630 &   -97.831\,832 &   -99.032\,131 \\
     &             & $E^{(1)}$           &    14.187\,481 &    -0.420\,871 &    -2.640\,975 &    -1.185\,613 &    -0.006\,999 \\
     &             & $E^{(2)}$           &    -0.385\,637 &     0.005\,979 &    -0.007\,828 &    -0.022\,403 &    -0.000\,164 \\
     &             & $E^{(3)}$           &    -0.002\,069 &    -0.000\,307 &     0.000\,083 &     0.000\,510 &    -0.000\,062 \\
     &             & $E^{(1)}_{\rm qed}$ &    -0.000\,927 &    -0.000\,730 &    -0.000\,758 &    -0.000\,786 &    -0.000\,753 \\
     &             & $E^{(2)}_{\rm qed}$ &     0.000\,165 &    -0.000\,038 &    -0.000\,011 &     0.000\,015 &    -0.000\,017 \\
     &             & Sum                 &   -99.040\,003 &   -99.040\,135 &   -99.040\,118 &   -99.040\,109 &   -99.040\,125 \\
\hline\\[-5pt]
  83 & $2s$        & $E^{(0)}$           &  -984.441\,516 &  -942.230\,603 &  -937.432\,166 &  -942.222\,843 &  -942.980\,768 \\
     &             & $E^{(1)}$           &    40.834\,128 &    -1.814\,029 &    -6.581\,926 &    -1.765\,359 &    -1.060\,437 \\
     &             & $E^{(2)}$           &    -0.431\,631 &     0.007\,841 &    -0.023\,024 &    -0.049\,251 &     0.004\,254 \\
     &             & $E^{(3)}$           &     0.001\,906 &    -0.000\,162 &     0.000\,252 &     0.000\,610 &    -0.000\,068 \\
     &             & $E^{(1)}_{\rm qed}$ &     0.015\,396 &     0.014\,348 &     0.014\,485 &     0.014\,629 &     0.014\,398 \\
     &             & $E^{(2)}_{\rm qed}$ &     0.005\,399 &     0.005\,946 &     0.005\,760 &     0.005\,643 &     0.005\,974 \\
     &             & Sum                 &  -944.016\,316 &  -944.016\,659 &  -944.016\,620 &  -944.016\,571 &  -944.016\,646 \\
     & $2p_{1/2}$  & $E^{(0)}$           &  -984.878\,756 &  -934.679\,963 &  -927.770\,428 &  -933.229\,693 &  -936.059\,025 \\
     &             & $E^{(1)}$           &    51.382\,561 &     0.376\,900 &    -6.488\,693 &    -0.973\,400 &     1.765\,487 \\
     &             & $E^{(2)}$           &    -0.819\,890 &    -0.002\,275 &    -0.048\,426 &    -0.106\,493 &    -0.012\,105 \\
     &             & $E^{(3)}$           &     0.006\,104 &    -0.000\,036 &     0.000\,775 &     0.001\,966 &     0.000\,098 \\
     &             & $E^{(1)}_{\rm qed}$ &     0.002\,191 &     0.000\,987 &     0.001\,094 &     0.001\,255 &     0.001\,070 \\
     &             & $E^{(2)}_{\rm qed}$ &     0.009\,687 &     0.005\,451 &     0.006\,839 &     0.007\,651 &     0.005\,575 \\
     &             & Sum                 &  -934.298\,103 &  -934.298\,936 &  -934.298\,840 &  -934.298\,713 &  -934.298\,900 \\
     & $2p_{3/2}$  & $E^{(0)}$           &  -881.829\,758 &  -839.911\,851 &  -833.217\,372 &  -837.510\,811 &  -840.767\,687 \\
     &             & $E^{(1)}$           &    41.873\,313 &    -0.544\,563 &    -7.218\,779 &    -2.903\,707 &     0.313\,613 \\
     &             & $E^{(2)}$           &    -0.497\,919 &     0.002\,689 &    -0.017\,588 &    -0.039\,424 &     0.000\,358 \\
     &             & $E^{(3)}$           &     0.001\,000 &    -0.000\,067 &     0.000\,171 &     0.000\,483 &    -0.000\,027 \\
     &             & $E^{(1)}_{\rm qed}$ &    -0.148\,479 &    -0.136\,186 &    -0.137\,893 &    -0.139\,618 &    -0.136\,915 \\
     &             & $E^{(2)}_{\rm qed}$ &     0.012\,714 &     0.000\,428 &     0.001\,946 &     0.003\,610 &     0.001\,115 \\
     &             & Sum                 &  -840.589\,129 &  -840.589\,550 &  -840.589\,515 &  -840.589\,468 &  -840.589\,542 \\
\end{tabular}
\end{ruledtabular}
\end{center}
\end{table*}

%
%
%
\begin{table}[htb]
\begin{center}
\caption{
Electron-structure part of the $2p_{1/2}$--$2s$ and $2p_{3/2}$--$2s$ transition energies (in eV),
compared with other calculations.
The uncertainties due to nuclear radii are not shown.
TW stands for this work.
\label{tab:elstruct:2}
}
\begin{ruledtabular}
\begin{tabular}{lw{2.9}w{2.9}l}
           \multicolumn{1}{c}{$Z$}
                &  \multicolumn{1}{c}{$2p_{1/2}$-$2s$}
                    &  \multicolumn{1}{c}{$2p_{3/2}$-$2s$}
                    &  \multicolumn{1}{c}{Ref.}
\\ \hline\\[-9pt]
 10   &    15.906\,94\,(15)    &       16.111\,08\,(15) & TW    \\
      &    15.906\,7\,(6)      &       16.110\,8\,(6)  &  \cite{kozhedub:10} \\
      &    15.906\,4\,(11)     &       16.110\,5\,(11) & \cite{yerokhin:07:lilike}\\
 15   &    25.885\,10\,(20)    &       27.273\,62\,(19) & TW   \\
      &    25.884\,8\,(3)      &       27.273\,2\,(3) &  \cite{kozhedub:10} \\
      &    25.885\,1\,(8)      &       27.273\,4\,(8) & \cite{yerokhin:07:lilike}\\
 20   &    36.163\,46\,(28)    &       41.215\,66\,(26)  & TW  \\
      &    36.163\,3\,(3)      &       41.215\,7\,(3) &  \cite{kozhedub:10} \\
      &    36.163\,4\,(4)      &       41.215\,5\,(4) & \cite{yerokhin:07:lilike}\\
 26   &    49.103\,45\,(27)    &       65.033\,89\,(21)  & TW  \\
      &    49.103\,0\,(3)      &       65.033\,9\,(3) &  \cite{kozhedub:10} \\
      &    49.102\,9\,(4)      &       65.033\,3\,(4) & \cite{yerokhin:07:lilike}\\
 30   &    58.213\,49\,(30)    &       87.793\,34\,(23)  & TW  \\
      &    58.213\,5\,(3)      &       87.793\,6\,(3) &  \cite{kozhedub:10} \\
      &    58.213\,0\,(4)      &       87.792\,6\,(4) & \cite{yerokhin:07:lilike}\\
 40   &    83.270\,76\,(37)    &       185.148\,10\,(42)  & TW  \\
      &    83.270\,6\,(11)     &       185.149\,0\,(11) &  \cite{kozhedub:10} \\
 50   &    112.744\,73\,(58)   &       379.033\,25\,(75)  & TW  \\
      &    112.744\,0\,(22)    &       379.032\,1\,(22) &  \cite{kozhedub:10} \\
      &    112.743\,3\,(16)    &       379.032\,3\,(21) & \cite{yerokhin:07:lilike}\\
 60   &    148.380\,84\,(95)   &       737.364\,6\,(12)   & TW \\
      &    148.381\,2\,(40)    &       737.364\,6\,(40) &  \cite{kozhedub:10} \\
 70   &    192.106\,9\,(16)    &       1359.567\,2\,(19)   & TW \\
	  &    192.104\,(10)       &       1359.565\,(10) &  \cite{kozhedub:10} \\
 83   &    264.433\,4\,(30)    &       2814.394\,9\,(32)   & TW \\
      &    264.430\,(16)       &       2814.391\,(16) &  \cite{kozhedub:10} \\
      &    264.427\,(28)       &       2814.392\,(28) & \cite{yerokhin:07:lilike}\\
 92   &    322.285\,7\,(48)    &       4498.738\,4\,(44)   & TW \\
      &    322.296\,(7)        &       4498.753\,(7) &  \cite{kozhedub:10} \\
\end{tabular}
\end{ruledtabular}
\end{center}
\end{table}

\subsection{Electron-structure QED}

\label{sec:structqed}

We now consider the QED corrections to the DCB energy that originate from the
electron-electron interaction only. They are referred to as the electron-structure QED
effects in the following. We perform a complete evaluation of
the one- and two-photon electron-structure QED effects, for different
screening potentials $U$.
Previously, QED calculations of electronic structure of Li-like ions
were carried out in
Refs.~\cite{yerokhin:00:prl,yerokhin:01:2ph,artemyev:03,yerokhin:07:lilike,kozhedub:10,sapirstein:11}.

In this section we summarize formulas for the one- and two-photon electron-structure QED
corrections. These formulas represent
the difference of the full QED expressions and the corresponding MBPT
corrections which are already included in the DCB energy.

The QED part of the one-photon exchange correction to the ionization energy of the valence
state $v$ is given by
\begin{align} \label{eq:1phot}
E^{(1)}_{\rm qed} &\, =
  \sum_{c} \sum_P (-1)^P  \I{Pv}{Pc}{v}{c}(\Delta_{Pcc}) - U_{vv} - E^{(1)}\,,
\end{align}
where $P$ is the permutation operator interchanging the one-electron states, $(PvPc) = (vc)$ or
$(cv)$, $(-1)^P$ is the sign of the permutation, $\Delta_{ab} = \vare_a -\vare_b$ is the
difference of one-electron energies, the summation over $c$ runs over the core electron states,
and $E^{(1)}$ is the one-photon MBPT correction given by Eq.~(\ref{eq:20:b}).
Furthermore, $I_{abcd}(\Delta) \equiv \lbr ab|I(\Delta)|cd\rbr$, where $I(\Delta)$ is the
full-QED operator of the electron-electron interaction. In the Feynman gauge,
\begin{equation}
  I(\omega) = \alpha\,
  \big( 1-\balpha_1 \cdot \balpha_2 \big)\,
  \frac{e^{i\sqrt{\omega^2+i0}\,x_{12}}}{x_{12}} \,,
\end{equation}
where
$x_{12} = |\bm{x}_{12}| = |\bm{x}_{1}-\bm{x}_{2}|$.

\begin{widetext}
The QED part of the two-photon exchange correction to the ionization energy of the valence
state of a Li-like atom is given by
\cite{yerokhin:01:2ph,kozhedub:10}
\begin{align}\label{eq:2ph}
E^{(2)}_{\rm qed} = &\ \sum_{c}\sum_{P}(-1)^P
    \left.\sum_{n_1n_2}\right.^{\!\prime}
 \frac{i}{2\pi} \intinf d\omega\,
    \Bigg[
     \frac{\I{Pc}{Pv}{n_1}{n_2}(\omega)\, \I{n_1}{n_2}{c}{v}(\omega-\Delta_{Pcc})}
     {(\vare_{Pc} -\omega -u\vare_{n_1})(\vare_{Pv} +\omega -u\vare_{n_2})}
  +   \frac{\I{Pc}{n_2}{n_1}{v}(\omega)\, \I{n_1}{Pv}{c}{n_2}(\omega-\Delta_{Pcc})}
     {(\vare_{Pc} -\omega -u\vare_{n_1})(\vare_v -\omega -u\vare_{n_2})}
     \Bigg]
       \nonumber \\ &
{} +   \sum_{PQ}(-1)^{P+Q}\,
    \left.\sum_{n}\right.^{\prime}
       \frac{\I{P2}{P3}{n}{Q3}(\Delta_{P3Q3})\, \I{P1}{n}{Q1}{Q2}(\Delta_{Q1P1})}
      {\vare_{Q1}+\vare_{Q2}-\vare_{P1}-\vare_{n}}
      + E_{\rm red}
       \nonumber \\ &
-  2\,\sum_{c} \sum_P (-1)^P \Big[ \I{Pv}{Pc}{\delta v}{c}(\Delta_{Pcc})
                               +\I{Pv}{Pc}{ v}{\delta c}(\Delta_{Pcc}) \Big]
+ \sum_{c} (U_{vv}-U_{cc})\, I^{\prime}_{cvvc}(\Delta_{vc}) + U_{v\,\delta v}
      - E^{(2)}\,,
\end{align}
where $P$ and $Q$ are the permutation operators, $u \equiv 1-i0$, the prime on the sum symbol
means that some terms are excluded from the summation (the excluded terms are ascribed to the
reducible part $E_{\rm red}$ and evaluated separately, see
Refs.~\cite{yerokhin:01:2ph,artemyev:03} for details),
$|\delta a\rbr = \sum_{n\neq a} U_{an}|n\rbr/(\vare_a-\vare_n)$,
and $I^{\prime}_{abcd}(\omega) = \lbr ab| \partial/(\partial \omega) I(\omega)|cd\rbr$.
\end{widetext}

In Eq.~(\ref{eq:2ph}), the first part on
the right-hand side is the irreducible two-electron contribution, the second part is the
irreducible three-electron contribution (with "1", "2", and "3" numerating the three electrons,
in arbitrary order), and the third part $\Delta E_{\rm red}$ is the reducible contribution
which is detailed out in Refs.~\cite{yerokhin:01:2ph,artemyev:03}.
The last line of Eq.~(\ref{eq:2ph}) contains terms induced by the screening potential $U$ and
and $E^{(2)}$ is the MBPT two-photon correction given by Eq.~(\ref{eq:21}).

Our present treatment of the electron-structure effects
extends the previous calculations by one of us reported in Ref.~\cite{yerokhin:07:lilike}.
While the general approach is similar, there are several differences. First,
the separation into the MBPT and QED parts is
different. In Ref.~\cite{yerokhin:07:lilike}, the MBPT part was defined to contain only
at most one Breit interaction and excluded parts induced by two and
three Breit interactions. In the present work we include these previously excluded parts into the definition
of the DCB energy, to ensure the compatibility between the MBPT and CI approaches.
The second difference is that in the present work we employ the MBPT formulas
from Ref.~\cite{blundell:87:adndt}, rather than those obtained in Ref.~\cite{yerokhin:07:lilike}.
The numerical results obtained with both sets of formulas agree well with each other (after the
contribution of two and three Breit interactions is separated out), but formulas of
Ref.~\cite{blundell:87:adndt} were found to be more numerically stable.

\subsection{Numerical results}
\label{sec:structnum}

We start with discussing our numerical results obtained for the DCB energies.
Table~\ref{tab:mbptci} presents a comparison of our values
calculated with the MBPT and CI methods,
for the LDF starting potential. The MBPT values include perturbative
corrections up to the three-photon exchange term, $E^{(3)}$.
The associated uncertainties were estimated as
the maximum deviation between the values obtained with three ``best''
screening potentials: KS, DS, and LDF.

Our CI calculations included single, double, and the dominant part of
triple excitations.  The partial-wave expansion was truncated
at $l = 9$ for calculations with the pure Coulomb interaction and at $l = 7$ for the Breit
interaction. The residual contribution from higher partial waves was estimated
by extrapolation in $1/l$. The contribution from triple excitations was found to
be small and was evaluated using a reduced basis set that included states with
$l \le 3$. As in our previous CI studies~\cite{yerokhin:08:pra,yerokhin:12:lilike},
the computations were performed using a large number (approximately 50) of different
basis sets. These sets varied in the number of partial waves included, the size of
the one-electron basis for each partial wave, the types of excitations considered,
and the inclusion or omission of the Breit interaction, etc.
The uncertainty of the CI energy was estimated by analyzing the convergence
behavior with respect to the systematic extension of the CI basis.

We observe that the results obtained with the MBPT and CI approaches are fully consistent within the estimated uncertainties.
In the high-$Z$ region, MBPT is preferable for two reasons: first, its calculations are considerably less time-consuming, and second, the MBPT framework is explicitly compatible with our QED treatment.
For low-$Z$ ions, however, the CI approach provides higher accuracy, owing to the missing four-photon
exchange contribution in the MBPT treatment. 
It is therefore advantageous to employ CI in this region.
The agreement between MBPT and CI results, as shown in Table~\ref{tab:mbptci}, ensures that the CI values can be safely combined with the QED treatment applied throughout the rest of this work.

We now turn to our calculations of the total electron-structure energies, which include both
the DCB and QED contributions. A detailed breakdown of the electron-structure part of the
ionization energies is presented in Table~\ref{tab:elstruct:1} for two representative ions: zinc and bismuth.
The table lists results obtained for the case of pure Coulomb starting potential and
four different choices of the screening potentials.
Contributions to the DCB energy listed in the table were calculated within the MBPT approach
up to third order of perturbation theory. The relativistic DCB treatment was
complemented by calculations of the one- and two-photon QED electron-structure corrections.
It can be clearly seen that the dependence of the total results on the choice
of the starting potential gradually diminishes as the number of perturbative-expansion
terms increases. We also observe that the convergence of the perturbation
expansion is much faster for the screening potentials than for the pure Coulomb potential.

We now need to obtain the final values for the electron-structure part of
energies and estimate the uncertainty due to the residual
electron correlation and QED effects.
As a central value we take the results obtained with the LDF potential
which is the closest to the Dirac-Fock potential.
In order to estimate the error bars of the central value, we add quadratically two
uncertainties: (i) the maximal deviation between values obtained with three ``best''
potentials (KS, DS, and LDF) and (ii) $2\, E^{(2)}_{\rm qed}/Z$.
The first uncertainty accounts for the residual electron correlation and negative-continuum
effects, whereas the second estimates the residual QED effects with exchange of three photons.

A separate procedure was used for evaluating transition energies of light ions with $Z \le 20$
and their uncertainties.
As demonstrated in Table~\ref{tab:mbptci}, for these ions the CI method provides more accurate
results than MBPT.
So, for $Z \le 20$
we used the CI values instead of the MBPT ones
and substituted the uncertainty (i) by the error estimate of the CI
values specified in Table~\ref{tab:mbptci}.

Table~\ref{tab:elstruct:2} shows a comparison of our numerical results for the
electron-structure part of transition energies with the previous calculations
\cite{kozhedub:10,yerokhin:07:lilike}. We find that all three calculations are
consistent with each other. Our numerical results, however, are more accurate.
This improvement was achieved (i) by combining together the MBPT and
CI methods for evaluating the DCB energies
and (ii) by a careful assessment of the uncertainty by analysing the
dependence of the obtained results on the starting screening potential.

\section{Radiative QED effects}

\label{sec:qed}

\subsection{One-electron QED}
\label{sec:qed1el}

In the one-electron approximation, the QED effects of the first order in $\alpha$ are the
self-energy and vacuum polarization.
The formal (unrenormalized) expression for the self-energy correction to the energy of the
valence electron state $v$ is
\begin{align}
E_{\rm se} = \bra{v} \Sigma(\vare_v)\ket{v} \,,
\end{align}
where the matrix element of the
one-loop self-energy operator $ \Sigma(\vare)$ is
defined by
\begin{align}
\lbr a | \Sigma(\vare)| b\rbr = \frac{i}{2\pi}\intinf d\omega \sum_n
  \frac{\I{a}{n}{n}{b}(\omega)}{\vare-\omega - u\vare_n}\,,
\end{align}
where the sum over $n$ is extended over the complete spectrum of the Dirac equation
and $u = 1 -i0$.

The unrenormalized expression for the vacuum-polarization correction is
given by the expectation value of the
vacuum-polarization potential $U_{\rm vp}$,
\begin{align}
E_{\rm vp} = \bra{v} U_{\rm vp}\ket{v} \,,
\end{align}
with
\begin{align}
U_{\rm vp}(\bfx) = \frac{\alpha}{2\pi i}
 \intinf d\omega \, \int d^3\bfy \, \frac{1}{|\bfx-\bfy|}\,{\rm Tr} \big[ G(\omega,\bfy,\bfy)\big]\,,
\end{align}
where $G(\omega) = (\omega-h_D)^{-1}$ is the Dirac-Coulomb Green function.

The one-electron QED corrections can be calculated both within the standard Furry picture
with the pure Coulomb starting potential and an extended Furry picture with a screening potential.
We here define $E_{\rm se}$ and $E_{\rm vp}$
specifically with respect to the pure Coulomb potential.
The deviations from these Coulomb-potential values due to the presence of a screening potential
are assigned to the QED screening effect, which is discussed in the following.

With this definition, the one-electron QED corrections
$E_{\rm se}$ and $E_{\rm vp}$ coincide with those for the hydrogenic $2s$ state,
tabulated in Ref.~\cite{yerokhin:15:Hlike}.
The same holds for the {\em two-loop} one-electron QED corrections,
which were taken from Ref.~\cite{yerokhin:15:Hlike} and the recent updates \cite{yerokhin:25:sese,volkov:25:tobe}.
Notably, our present uncertainties in the two-loop QED correction are reduced compared to those
in Ref.~\cite{yerokhin:15:Hlike} by about 50\%,
owing to the recent calculation of the two-loop vacuum polarization
\cite{volkov:25:tobe}, which was previously one of the main sources of uncertainty.

\subsection{QED screening}
\label{sec:qedscr}

The presence of core electrons modifies the self-energy and vacuum polarization of the valence electron,
an effect known as {\em screening}.
We describe the screening effect on the self-energy and vacuum polarization by expanding in the number of photons exchanged between the valence and core electrons,
\begin{align} \label{eq:qedscr}
E_{\rm sescr} = &\, E_{\rm sescr}^{(0)} + E_{\rm sescr}^{(1)} + E_{\rm sescr}^{(2)} + \ldots\,,
 \nonumber \\
E_{\rm vpscr} = &\, E_{\rm vpscr}^{(0)} + E_{\rm vpscr}^{(1)} + E_{\rm vpscr}^{(2)} + \ldots\,.
\end{align}
In the present work we calculate rigorously the first two terms of the above expansion. The terms
with two photon exchanges, $E^{(2)}$, will be calculated approximately by using the model QED
operator.

The leading terms in the expansion (\ref{eq:qedscr}) contain zero exchanged photons. They are
obtained as a difference between the one-electron self-energy and vacuum-polarization corrections
calculated with the screening potential $U$ and with the pure Coulomb potential ($U = 0$),
\begin{align} \label{eq:qedsc2}
E_{\rm sescr}^{(0)}  = E_{\rm se}(U) - E_{\rm se}(U=0)\,,
 \nonumber \\
E_{\rm vpscr}^{(0)}  = E_{\rm vp}(U) - E_{\rm vp}(U=0)\,.
\end{align}

%
%
%
\begin{table*}[htb]
\begin{center}
\caption{
Self-energy screening correction calculated for different
starting potentials for the $2s$, $2p_{1/2}$, and $2p_{3/2}$
states of Li-like ions.
Units are $\delta E/[\alpha^2(\Za)^3\,mc^2]$.
\label{tab:sescr}
}
\begin{ruledtabular}
\begin{tabular}{llcw{2.7}w{2.7}w{2.7}w{2.7}w{2.10}}
 $Z$ & State & Term
              &  \multicolumn{1}{c}{Coul}
                &  \multicolumn{1}{c}{CH}
                    &  \multicolumn{1}{c}{KS}
                        &  \multicolumn{1}{c}{DS}
                            &  \multicolumn{1}{c}{Final}
\\ \hline\\[-9pt]
%
 10 & $2s$       &  $E^{(0)}_{\rm sescr}$  &                     & -0.581\,91          & -0.564\,16          & -0.499\,44          &   \\
    &            &  $E^{(1)}_{\rm sescr}$  & -0.756\,50\,(6)     & -0.109\,14\,(2)     & -0.125\,78\,(2)     & -0.196\,93\,(2)     &   \\
    &            &  $E^{(2)}_{\rm sescr}$  &  0.058\,18          &  0.008\,38          &  0.005\,26          &  0.010\,04          &   \\
    &            &  Sum                    & -0.698\,31\,(6)     & -0.682\,67\,(2)     & -0.684\,68\,(2)     & -0.686\,34\,(2)     & -0.6847\,(55) \\[2pt]
%
    & $2p_{1/2}$ &  $E^{(0)}_{\rm sescr}$  &                     &  0.011\,86          &  0.006\,95          &  0.004\,01          &   \\
    &            &  $E^{(1)}_{\rm sescr}$  & -0.164\,62\,(7)     & -0.123\,52\,(4)     & -0.117\,41\,(2)     & -0.119\,74\,(3)     &   \\
    &            &  $E^{(2)}_{\rm sescr}$  &  0.057\,02          &  0.008\,41          &  0.006\,20          &  0.012\,29          &   \\
    &            &  Sum                    & -0.107\,60\,(7)     & -0.103\,26\,(4)     & -0.104\,26\,(2)     & -0.103\,44\,(3)     & -0.1043\,(19) \\[2pt]
%
    & $2p_{3/2}$ &  $E^{(0)}_{\rm sescr}$  &                     & -0.037\,69          & -0.038\,00          & -0.035\,09          &   \\
    &            &  $E^{(1)}_{\rm sescr}$  & -0.226\,51\,(6)     & -0.123\,85\,(4)     & -0.122\,82\,(2)     & -0.132\,18\,(3)     &   \\
    &            &  $E^{(2)}_{\rm sescr}$  &  0.063\,99          &  0.008\,56          &  0.006\,39          &  0.012\,92          &   \\
    &            &  Sum                    & -0.162\,52\,(6)     & -0.152\,98\,(4)     & -0.154\,44\,(2)     & -0.154\,35\,(3)     & -0.1544\,(23) \\[2pt]
%
 20 & $2s$       &  $E^{(0)}_{\rm sescr}$  &                     & -0.440\,03          & -0.420\,03          & -0.367\,23          &   \\
    &            &  $E^{(1)}_{\rm sescr}$  & -0.546\,66\,(3)     & -0.082\,10          & -0.102\,26          & -0.157\,57          &   \\
    &            &  $E^{(2)}_{\rm sescr}$  &  0.021\,20          &  0.001\,82          &  0.001\,68          &  0.003\,26          &   \\
    &            &  Sum                    & -0.525\,46\,(3)     & -0.520\,31          & -0.520\,61          & -0.521\,53          & -0.5206\,(18) \\[2pt]
%
    & $2p_{1/2}$ &  $E^{(0)}_{\rm sescr}$  &                     &  0.007\,44          &  0.002\,93          &  0.000\,89          &   \\
    &            &  $E^{(1)}_{\rm sescr}$  & -0.116\,37\,(3)     & -0.104\,19          & -0.099\,57          & -0.099\,59          &   \\
    &            &  $E^{(2)}_{\rm sescr}$  &  0.021\,17          &  0.003\,06          &  0.002\,68          &  0.004\,90          &   \\
    &            &  Sum                    & -0.095\,19\,(3)     & -0.093\,69          & -0.093\,95          & -0.093\,80          & -0.0940\,(7) \\[2pt]
%
    & $2p_{3/2}$ &  $E^{(0)}_{\rm sescr}$  &                     & -0.045\,05          & -0.043\,68          & -0.038\,75          &   \\
    &            &  $E^{(1)}_{\rm sescr}$  & -0.172\,41\,(2)     & -0.102\,66          & -0.104\,14\,(1)     & -0.111\,65\,(1)     &   \\
    &            &  $E^{(2)}_{\rm sescr}$  &  0.023\,69          &  0.003\,18          &  0.002\,73          &  0.005\,12          &   \\
    &            &  Sum                    & -0.148\,72\,(2)     & -0.144\,53          & -0.145\,09\,(1)     & -0.145\,28\,(1)     & -0.1451\,(12) \\[2pt]
%
 50 & $2s$       &  $E^{(0)}_{\rm sescr}$  &                     & -0.300\,65\,(1)     & -0.280\,95\,(1)     & -0.242\,21\,(1)     &   \\
    &            &  $E^{(1)}_{\rm sescr}$  & -0.369\,26\,(1)     & -0.060\,56          & -0.080\,64          & -0.120\,28          &   \\
    &            &  $E^{(2)}_{\rm sescr}$  &  0.006\,81          & -0.000\,21          &  0.000\,37          &  0.000\,94          &   \\
    &            &  Sum                    & -0.362\,45\,(1)     & -0.361\,43          & -0.361\,22          & -0.361\,55          & -0.361\,22\,(50) \\[2pt]
%
    & $2p_{1/2}$ &  $E^{(0)}_{\rm sescr}$  &                     & -0.018\,27          & -0.018\,34          & -0.016\,25          &   \\
    &            &  $E^{(1)}_{\rm sescr}$  & -0.101\,57          & -0.075\,60          & -0.075\,84          & -0.078\,82          &   \\
    &            &  $E^{(2)}_{\rm sescr}$  &  0.007\,62          &  0.000\,68          &  0.000\,88          &  0.001\,71          &   \\
    &            &  Sum                    & -0.093\,95          & -0.093\,19          & -0.093\,30          & -0.093\,36          & -0.093\,30\,(27) \\[2pt]
%
    & $2p_{3/2}$ &  $E^{(0)}_{\rm sescr}$  &                     & -0.059\,77          & -0.055\,06          & -0.047\,31          &   \\
    &            &  $E^{(1)}_{\rm sescr}$  & -0.128\,78\,(6)     & -0.061\,73          & -0.066\,59          & -0.075\,09          &   \\
    &            &  $E^{(2)}_{\rm sescr}$  &  0.006\,73          &  0.000\,76          &  0.000\,80          &  0.001\,49          &   \\
    &            &  Sum                    & -0.122\,05\,(6)     & -0.120\,75          & -0.120\,85          & -0.120\,92          & -0.120\,85\,(26) \\[2pt]
%
 83 & $2s$       &  $E^{(0)}_{\rm sescr}$  &                     & -0.308\,20          & -0.279\,23          & -0.236\,13          &   \\
    &            &  $E^{(1)}_{\rm sescr}$  & -0.389\,59          & -0.074\,58          & -0.104\,18          & -0.148\,19          &   \\
    &            &  $E^{(2)}_{\rm sescr}$  &  0.006\,34          & -0.000\,08          &  0.000\,82          &  0.001\,50          &   \\
    &            &  Sum                    & -0.383\,25          & -0.382\,87          & -0.382\,58          & -0.382\,82          & -0.382\,58\,(43) \\[2pt]
%
    & $2p_{1/2}$ &  $E^{(0)}_{\rm sescr}$  &                     & -0.083\,05          & -0.072\,80          & -0.060\,54          &   \\
    &            &  $E^{(1)}_{\rm sescr}$  & -0.185\,59          & -0.093\,05          & -0.104\,18          & -0.117\,70          &   \\
    &            &  $E^{(2)}_{\rm sescr}$  &  0.008\,61          &  0.000\,95          &  0.001\,46          &  0.002\,40          &   \\
    &            &  Sum                    & -0.176\,97          & -0.175\,14          & -0.175\,52          & -0.175\,83          & -0.1755\,(10) \\[2pt]
%
    & $2p_{3/2}$ &  $E^{(0)}_{\rm sescr}$  &                     & -0.079\,86          & -0.072\,23          & -0.061\,69          &   \\
    &            &  $E^{(1)}_{\rm sescr}$  & -0.128\,33\,(30)    & -0.044\,13          & -0.051\,89\,(4)     & -0.062\,85          &   \\
    &            &  $E^{(2)}_{\rm sescr}$  &  0.004\,33          &  0.000\,26          &  0.000\,47          &  0.000\,95          &   \\
    &            &  Sum                    & -0.124\,01\,(30)    & -0.123\,73          & -0.123\,64\,(4)     & -0.123\,60          & -0.123\,64\,(21) \\
\end{tabular}
\end{ruledtabular}
\end{center}
\end{table*}

%
%
%
\begin{table}[htb]
\begin{center}
\caption{
The QED screening correction for transition energies of Li-like ions,
in eV.
\label{tab:qedscr}
}
\begin{ruledtabular}
\begin{tabular}{lw{2.9}w{2.9}l}
 $Z$ &   \multicolumn{1}{c}{$2p_{1/2}$--$2s$}
                &  \multicolumn{1}{c}{$2p_{3/2}$--$2s$}
                &  \multicolumn{1}{c}{Ref.}
\\ \hline\\[-9pt]
 10    &  0.00585\,(6)          &  0.00531\,(4) & TW   \\
       &  0.0058\,(2)           &  0.0053\,(2) & \cite{kozhedub:10} \\
 12    &  0.00934\,(7)          &  0.00839\,(4) & TW    \\
       &  0.0094\,(2)           &  0.0085\,(3) & \cite{kozhedub:10} \\
 14    &  0.01382\,(9)          &  0.01231\,(5)  & TW   \\
       &  0.0138\,(3)           &  0.0123\,(3) & \cite{kozhedub:10} \\
 18    &  0.02600\,(13)         &  0.02279\,(8)   & TW  \\
       &  0.0260\,(4)           &  0.0228\,(5) & \cite{kozhedub:10} \\
 20    &  0.03380\,(15)         &  0.02943\,(10)  & TW   \\
       &  0.0338\,(5)           &  0.0294\,(5) & \cite{kozhedub:10} \\
 26    &  0.06455\,(23)         &  0.05534\,(16)  & TW   \\
       &  0.0646\,(8)           &  0.0554\,(8) & \cite{kozhedub:10} \\
 30    &  0.09160\,(29)         &  0.07803\,(20)  & TW   \\
       &  0.0917\,(10)          &  0.0782\,(11) & \cite{kozhedub:10} \\
 40    &  0.18389\,(44)         &  0.15610\,(35)  & TW   \\
       &  0.1840\,(17)          &  0.1561\,(18) & \cite{kozhedub:10} \\
 50    &  0.31411\,(67)         &  0.27125\,(64)  & TW   \\
       &  0.3141\,(26)          &  0.2713\,(27) & \cite{kozhedub:10} \\
 60    &  0.4843\,(14)          &  0.43617\,(97)  & TW   \\
       &  0.4841\,(42)          &  0.4361\,(38) & \cite{kozhedub:10} \\
 70    &  0.6927\,(26)          &  0.6715\,(14)   & TW  \\
       &  0.692\,(7)            &  0.672\,(5) & \cite{kozhedub:10} \\
 80    &  0.9282\,(54)          &  1.0106\,(26)   & TW  \\
       &  0.928\,(11)           &  1.014\,(8) & \cite{kozhedub:10} \\
 83    &  1.0000\,(71)          &  1.1398\,(31)  & TW   \\
       &  1.000\,(13)           &  1.141\,(9) & \cite{kozhedub:10} \\
 90    &  1.155\,(14)           &  1.5050\,(46)  & TW   \\
       &  1.153\,(17)           &  1.512\,(11) & \cite{kozhedub:10} \\
 92    &  1.193\,(16)           &  1.6285\,(53)   & TW  \\
       &  1.190\,(19)           &  1.637\,(11)& \cite{kozhedub:10} \\
100    &  1.277\,(36)           &  2.229\,(10)   & TW  \\
\end{tabular}
\end{ruledtabular}
\end{center}
\end{table}

The second terms in the expansion (\ref{eq:qedscr}) contain one exchanged photon between the valence
electron and the core.
The derivation of the general formulas for the screened self-energy correction for Li-like ions
was reported in
Ref.~\cite{yerokhin:99:sescr}, see also Ref.~\cite{kozhedub:10}.
We here rearrange the formulas into a form optimal
for a numerical evaluation.
The screened self-energy correction is conveniently represented as a sum of the perturbed-orbital (po),
reducible (red), and vertex (ver) contributions,
\begin{align}\label{eq:se:0}
\Delta E_{\rm sescr}^{(1)} = \Delta E_{\rm sescr, po} + \Delta E_{\rm sescr, red} + \Delta E_{\rm sescr, ver}\,.
\end{align}
The perturbed-orbital contribution is expressed in terms of diagonal and non-diagonal
matrix elements of the one-loop self-energy operator
$ \Sigma(\vare)$,
\begin{widetext}
\begin{align} \label{eq:se:1}
\Delta E_{\rm sescr, po} = &\ \sum_{PQ}(-1)^{P+Q}
\bigg[
2 \sum_{n \neq Pv}
  \lbr Pv | \Sigma(\vare_{Pv})| n\rbr\,
  \frac{\I{n}{Pc}{Qv}{Qc}(\Delta_{QcPc})
   - \delta_{Pv,v}\,\delta_{Qv,v}\,U_{nv}
  }{\vare_{Pv}-\vare_{n}}
\nonumber \\ &
 + \lbr Pv | \Sigma(\vare_{Pv})| Pv\rbr\,
  I^{\prime}_{PvPc\,QvQc}(\Delta_{QcPc})
\bigg]
\,,
\end{align}
where $P$ and $Q$ are the permutation operators, $(PvPc) = (vc), (cv)$, $(QvQc) = (vc), (cv)$,
and $\delta_{ik}$ is the Kronecker symbol.
The reducible part of the screened self-energy correction contains the derivative of the self-energy operator
over the energy argument and is given by
\begin{align} \label{eq:se:2}
\Delta E_{\rm sescr, red} = &\ \sum_{PQ}(-1)^{P+Q}
 \lbr Pv | \Sigma{}'(\vare_{Pv})| Pv\rbr\,
 \Big[
  \I{Pv}{Pc}{Qv}{Qc}(\Delta_{QcPc})
- \delta_{Pv,v}\, \delta_{Qv,v}\,U_{vv}
\Big]
 \,.
\end{align}
The vertex  part of the screened self-energy correction is
\begin{align} \label{eq:se:4}
\Delta E_{\rm sescr, ver} = &\ \sum_{PQ}(-1)^{P+Q}
\frac{i}{2\pi}
  \intinf d\omega \,
\sum_{n_1n_2}
\frac{\I{Pv}{n_2}{n_1}{Qv}(\omega)\,
 \Big[\I{n_1}{Pc}{n_2}{Qc}(\Delta_{QcPc})
   - \delta_{Pv,v}\,\delta_{Qv,v}\,U_{n_1n_2}
 \Big]}
  {(\vare_{Pv}-\omega-u\,\vare_{n_1})(\vare_{Qv}-\omega-u\,\vare_{n_2})}
  \,.
\end{align}
%
The screened vacuum-polarization correction \cite{artemyev:05:pra} is conveniently represented as a sum of the
perturbed-orbital (po), and the perturbed photon-propagator (ph) contributions,
\begin{align}\label{eq:vp:0}
\Delta E_{\rm vpscr}^{(1)} = \Delta E_{\rm vpscr, po} + \Delta E_{\rm vpscr, ph}\,.
\end{align}
The perturbed-orbital contribution is analogous to that for the screened self-energy and is expressed in terms of
matrix elements of the one-loop vacuum-polarization potential,
\begin{align} \label{eq:vp:1}
\Delta E_{\rm vpscr, po} = &\ \sum_{PQ}(-1)^{P+Q}
\bigg[
2 \sum_{n \neq Pv}
  \lbr Pv | U_{\rm vp}| n\rbr\,
  \frac{
  \I{n}{Pc}{Qv}{Qc}(\Delta_{QcPc})
   - \delta_{Pv,v}\,\delta_{Qv,v}\,U_{nv}
  }{\vare_{Pv}-\vare_{n}}
 + \lbr Pv | U_{\rm vp}| Pv\rbr\,
  I^{\prime}_{PvPc\,QvQc}(\Delta_{QcPc})
\bigg] \,.
\end{align}
The remaining part of the screened vacuum-polarization is given by the correction to the photon
propagator,
\begin{align} \label{eq:vp:2}
\Delta E_{\rm vpscr, ph} = &\ \sum_{PQ}(-1)^{P+Q}
  \lbr Pv Pc|I_{\rm vp}(\Delta_{QcPc})|QvQc\rbr
\nonumber \\ &
  -
\frac{\alpha}{2\pi i}
 \intinf d\omega \,
 \int d^3\bfx\,d^3\bfy\,d^3\bfz\,
 \psi_v^{\dag}(\bfx) \frac{1}{|\bfx-\bfy|}\,\psi_v(\bfx)\,
   {\rm Tr} \big[ G(\omega,\bfy,\bfz)\,U(z)\,G(\omega,\bfz,\bfy)\big]
 \,,
\end{align}
where
$I_{\rm vp}$ is the radiatively corrected electron-electron interaction operator,
\begin{align}
I_{\rm vp}(\delta,\bfx,\bfy) = \frac{\alpha^2}{2\pi i}
 \intinf d\omega \, \int d^3\bfz_1d^3\bfz_2 \,\alpha_{\mu} \,
  D^{\mu\nu}(\delta,\bfx,\bfz_1)\,
 {\rm Tr} \Big[
  \alpha_{\nu}\,G(\omega-\delta/2,\bfz_1,\bfz_2)\,
  \alpha_{\rho}\,G(\omega+\delta/2,\bfz_2,\bfz_1)
  \Big]
  D^{\rho\sigma}(\delta,\bfz_2,\bfy)\,
\alpha_{\sigma}\,
.
\end{align}
\end{widetext}

Our numerical evaluation of the screened self-energy correction is based on
the use of the Dirac Green function, computed for the case of a
general screening potential by the numerical procedure outlined in Ref.~\cite{yerokhin:11:fns}.
The perturbed-orbital contribution $\Delta E_{\rm po}$ is expressed in
terms of matrix elements of the one-loop self-energy operator
$\Sigma(\vare)$, which are computed by numerical methods described in detail in
Refs.~\cite{yerokhin:99:pra,yerokhin:05:se,yerokhin:25:se}. The general scheme of evaluation of the reducible
and vertex corrections was developed in Ref.~\cite{yerokhin:99:sescr}.
In the present work we adopt the partial-wave convergence-acceleration scheme by
Sapirstein and Cheng \cite{sapirstein:23} for the computation of the vertex
part of the self-energy screening correction, in the same manner as was recently done
in Ref.~\cite{malyshev:24}.
The partial-wave convergence acceleration method enabled a significant
improvement in the numerical accuracy of our results compared to previous calculations,
which is particularly relevant for low values of $Z$.

In our calculation of the perturbed photon-propagator part of the screening vacuum polarization
correction, we included only the contribution of the free-electron propagators in
the vacuum-polarization loop. The remaining part, known as the light-by-light
scattering contribution, is very small for Li-like ions
\cite{artemyev:99,artemyev:priv:21}, and its omission does not
affect the uncertainty of the total QED contribution.

We now turn to the evaluation of the $1/Z^2$ screening QED effect,
which corresponds to the screening of the self-energy and vacuum polarization
with two photons exchanged between the valence and core electrons. A rigorous
QED calculation of this contribution is currently not feasible, so we treat
it approximately using the model QED operator \cite{shabaev:13:qedmod}, as
implemented in the QEDMOD package \cite{shabaev:14:qedmod}.
Since the model QED potential $V_{\rm mqed}$ is a one-body operator, we obtain the
$1/Z^2$ correction induced by it as a first-order (in $V_{\rm mqed}$) perturbation of the
two-photon MBPT correction $E^{(2)}$ given by Eq.~(\ref{eq:21}).
Specifically, we perturb all single-electron energies and wave functions in Eq.~(\ref{eq:21})
as
\begin{align}\label{eq:24}
\vare_i \to \vare_i + \bra{i} V_{\rm mqed}\ket{i} \,,\ \ \
   \ket{i} \to  \ket{i}  + \sum_{k\neq i}\frac{\ket{k}\,\bra{k}V_{\rm mqed}\ket{i}}
 {\vare_i-\vare_k}\,,
\end{align}
and after that we pick up the linear in $V_{\rm mqed}$ contribution.
The summation over $k$ in Eq.~(\ref{eq:24}) runs over the spectrum of Dirac one-electron states.
The corresponding formulas are simple but rather lengthy and can be easily worked out
along the lines described in Ref.~\cite{yerokhin:07:lilike} for
other MBPT corrections.

The breakdown of our numerical calculations of the self-energy screening corrections
is presented in Table~\ref{tab:sescr}.
We carried out our calculations using three different screening potentials: CH, KS, and DS.
The LDF potential was omitted, as we were unable to achieve the desired level of numerical accuracy for this
potential.

We observe that the two-photon screening correction, which was previously
omitted in calculations of Li-like ions, yields a sizeable contribution, particularly for low-$Z$ ions.
Even with the inclusion of this correction, the results exhibit a significant dependence on
the choice of the starting potential.
Therefore, we need to select a final value carefully and provide a well-justified estimate of
its uncertainty.

As a central value of the self-energy screening correction we take the result obtained with the KS potential.
Its error bars are estimated by adding quadratically two uncertainties: (i) the maximal difference
between the values obtained with the three screening potentials,
multiplied by a conservative factor of 1.5
and (ii) the value of $E^{(2)}_{\rm sescr}$
scaled by the relative deviation of the QEDMOD approximation applied to $E^{(1)}_{\rm sescr}$ from
its exact value, multiplied by a conservative factor of 2. The resulting final values of the
self-energy screening correction are listed in the last column of Table~\ref{tab:sescr}.

Table~\ref{tab:qedscr} presents a comparison of our numerical results for the
QED screening contribution to the transition energies of Li-like ions with the previous calculation
by Kozhedub {\em et al.} \cite{kozhedub:10}.
We observe excellent agreement in all cases, with typical deviations significantly smaller than the uncertainties estimated in Ref.\cite{kozhedub:10}.
Our results are several times more accurate than those of Ref.~\cite{kozhedub:10},
owing to the inclusion of the two-photon screening contribution and a more detailed estimation of the uncertainties.

In the present work, we also include an estimate of the {\em two-loop} QED screening effect.
For the $2s$ state, the screening contribution is estimated by multiplying the one-electron two-loop QED
correction by the ratio of the one-loop QED screening correction to the corresponding hydrogenic contribution,
with a 50\% uncertainty assigned to the result.
For the $2p$ states, the two-loop QED screening contribution is assumed to be zero, and the same
uncertainty as for the $2s$ state is assigned.

\section{Nuclear recoil}
\label{sec:rec}

Within the Breit approximation, the relativistic nuclear recoil
effect can be described by the
relativistic operator
\cite{shabaev:85,shabaev:88,shabaev:98:rectheo}
\begin{eqnarray}\label{eq:rrec}
H_{\rm rrec} = \frac{m}{2M}\, \sum_{ij}\biggl[ \bfp_i\cdot\bfp_j -\frac{Z\alpha}{r_i} \left( \balpha_i
+ \frac{(\balpha_i\cdot\bfr_i)\,\bfr_i}{r_i^2}\right) \cdot\bfp_j \biggr]\,,
\end{eqnarray}
where $M$ is the nuclear mass,
$\bfp$ is the momentum operator, and $i$ and $j$ numerate the electrons.
Previous calculations of the relativistic recoil effect with operator $H_{\rm rrec}$
were reported for Li-like ions
in Refs.~\cite{kozhedub:10,zubova:14}.

In this work we use the CI method for computing the relativistic part of the
nuclear recoil correction,
by adding the operator $H_{\rm rrec}$ to the DCB Hamiltonian
and comparing the results obtained with and without
the $H_{\rm rrec}$ addition.

The Breit approximation describes the leading nuclear recoil correction of
order $(m/M)(\Za)^2$ and $(m/M)(\Za)^4$. This approximation becomes
not adequate for high-$Z$ ions, where the higher-order corrections
become prominent. These higher-order recoil corrections
can be obtained only within the full-QED treatment.

To the leading order in $1/Z$ and to all orders in $\Za$, the nuclear recoil contribution
to the ionization energy of the valence state $v$ of an
atom with one valence electron beyond the closed shell(s)
is given by
\cite{shabaev:98:rectheo}
\begin{align}\label{eq:rec}
E_{\rm rec} =&\,\frac{m}{M} \frac{i}{2\pi}\, \int_{-\infty}^{\infty}d\omega\,
\sum_n
 \frac1{\vare_v + \omega - \vare_n + i\eta 0}
\nonumber \\ & \times
 \bra{v} \bfp - {\bm D}(\omega) \ket{n}
   \bra{n} \bfp - {\bm D}(\omega) \ket{v}
   \,,
\end{align}
where the vector
${\bm D}(\omega)$ is
connected with the transverse part of the photon propagator in the Coulomb gauge $D_C^{ij}$ as
$$
D^j(\omega) = -4\pi Z\alpha \, \alpha^i \, D_C^{ij}(\omega,\vec{r})\,,
$$
and $\eta = \mbox{\rm sign}(\vare_n -\vare_c - \delta)$, with $\delta$ being small and positive
and $\vare_c$ being the Dirac energy of the outermost closed shell (in our case, the $1s$ state).
Note that Eq.~(\ref{eq:rec}) differs from the analogous expression for the hydrogen-like atom
only by the sign of the imaginary addition $i0$ for the core intermediate states.

We define the QED part of the nuclear recoil correction by subtracting from $E_{\rm rec}$
its Breit-approximation limit,
\begin{align}
E_{\rm rec,qed} = E_{\rm rec} - E_{\rm rec,Br}\,,
\end{align}
where
\begin{align}
E_{\rm rec,Br} = &\ \frac{m}{2M}\bra{v} \Big[\bfp^2 - \bfp\cdot{\bm D}(0) - {\bm D}(0)\cdot\bfp \Big] \ket{v}
\nonumber \\ &
- \frac{m}M\,\sum_c \bra{v} \bfp\ket{c}\bra{c} \bfp \ket{v}
\nonumber \\ &
+\frac{2m}M\, \sum_c \bra{v} \bfp\ket{c}\bra{c} {\bm D}(0) \ket{v}\,.
\end{align}

Previous calculations of the QED recoil corrections for Li-like ions have been
carried out to the leading order in the $1/Z$ expansion
in Refs.~\cite{artemyev:95:pra,artemyev:95:jpb,kozhedub:10}.
An extension of the QED treatment to the next-to-leading order in $1/Z$
has been reported by Malyshev et al.~\cite{malyshev:19:rec,malyshev:20:rec}.
In the present work we perform calculations of the QED recoil correction
to the leading order in $1/Z$, following the numerical procedure
detailed out in Ref.~\cite{yerokhin:23:recoil}.
Higher-order QED recoil corrections in $1/Z$ are estimated by multiplying 
the leading-order contribution by a factor of $5/Z$, which approximately 
corresponds to the ratio of the screening QED correction to the one-electron 
radiative QED shift.
Since the resulting uncertainty is relatively small compared with other 
theoretical uncertainties, we do not pursue the calculation of the 
next-to-leading QED effects as it was done
in Refs.~\cite{malyshev:19:rec,malyshev:20:rec}.

Numerical results of our calculations of the nuclear recoil correction are
presented in Table~\ref{tab:recoil}, in comparison with the
data reported in Ref.~\cite{zubova:14}.
We observe a good agreement between the two calculations. A small deviation
for large $Z$ is probably due to different treatments of the finite nuclear size
effect in the QED recoil part. In our work, we use the finite-size
photon propagator derived in Ref.~\cite{pachucki:23:prl},
whereas  Ref.~\cite{zubova:14} employed the standard point-nucleus
photon propagator. Larger uncertainties of our values are due to the
more conservative uncertainty estimation in this work.

%
%
%
\begin{table}[htb]
\begin{center}
\caption{
Nuclear recoil correction to the transition energies of Li-like ions. Units
are $\delta E/[(\Za)^2 (m/M)\,mc^2]$.
\label{tab:recoil}
}
\begin{ruledtabular}
\begin{tabular}{lw{2.9}w{2.9}l}
 $Z$ &   \multicolumn{1}{c}{$2p_{1/2}$--$2s$}
                &  \multicolumn{1}{c}{$2p_{3/2}$--$2s$}
                &  \multicolumn{1}{c}{Ref.}
\\ \hline\\[-9pt]
 10 &  -0.056\,65\,(4)          & -0.056\,75\,(4) & TW\\
    &  -0.056\,65\,(1)          & -0.056\,75\,(1) &  \cite{zubova:14} \\
 14 &  -0.062\,63\,(7)          & -0.062\,88\,(7) & TW\\
    &  -0.062\,62\,(3)          & -0.062\,88\,(3) &  \cite{zubova:14} \\
 18 &  -0.065\,98\,(10)         & -0.066\,46\,(10)& TW \\
    &  -0.065\,98\,(4)          & -0.066\,46\,(4) &  \cite{zubova:14} \\
 22 &  -0.068\,15\,(15)         & -0.068\,93\,(15)& TW \\
    &  -0.068\,15\,(7)          & -0.068\,92\,(6) &  \cite{zubova:14} \\
 26 &  -0.069\,70\,(21)         & -0.070\,86\,(21)& TW \\
    &  -0.069\,71\,(9)          & -0.070\,86\,(9) &  \cite{zubova:14} \\
 30 &  -0.070\,94\,(27)         & -0.072\,57\,(27)& TW \\
    &  -0.070\,96\,(12)         & -0.072\,56\,(12) &  \cite{zubova:14} \\
 36 &  -0.072\,51\,(39)         & -0.075\,01\,(39) & TW\\
    &  -0.072\,51\,(17)         & -0.074\,97\,(17) &  \cite{zubova:14} \\
 42 &  -0.074\,03\,(54)         & -0.077\,64\,(53) & TW\\
    &  -0.074\,10\,(24)         & -0.077\,57\,(22) &  \cite{zubova:14} \\
 54 &  -0.077\,77\,(93)         & -0.084\,52\,(93) & TW\\
    &  -0.077\,91\,(38)         & -0.084\,37\,(38) &  \cite{zubova:14} \\
 60 &  -0.080\,4\,(12)          & -0.089\,3\,(12) & TW\\
    &  -0.080\,65\,(54)         & -0.089\,19\,(54) &  \cite{zubova:14} \\
 70 &  -0.086\,7\,(18)          & -0.100\,7\,(18) & TW\\
    &  -0.087\,30\,(85)         & -0.100\,59\,(79) &  \cite{zubova:14} \\
 80 &  -0.096\,9\,(28)          & -0.118\,7\,(28) & TW\\
    &  -0.098\,4\,(13)          & -0.119\,1\,(12) &  \cite{zubova:14} \\
 83 &  -0.101\,1\,(32)          & -0.126\,1\,(32) & TW\\
    &  -0.103\,5\,(14)          & -0.127\,4\,(14) &  \cite{zubova:14} \\
 90 &  -0.113\,7\,(44)          & -0.148\,2\,(43) & TW\\
    &  -0.117\,7\,(19)          & -0.151\,0\,(18) &  \cite{zubova:14} \\
 92 &  -0.118\,1\,(49)          & -0.156\,2\,(48) & TW\\
    &  -0.122\,2\,(21)          & -0.159\,1\,(20) &  \cite{zubova:14} \\
%
\end{tabular}
\end{ruledtabular}
\end{center}
\end{table}

\section{Nuclear effects}
\label{sec:nucl}

The dominant part of the nuclear contributions arises from the finite nuclear size (fns) effect,
which is already included into the DCB energies and QED corrections calculated for a
finite nuclear charge distribution.
The fns effect was incorporated in the DCB and QED calculations by
using the standard two-parameter Fermi model
(see, for example, Ref.~\cite{yerokhin:15:Hlike} for details).
The nuclear charge radii were taken from the compilation by Angeli and Marinova \cite{angeli:13}.
In rare
cases where no data for the nuclear radius were available, the
standard empirical formula \cite{johnson:85} was employed (in fermi)
$$
R = 0.836\, A^{1/3} + 0.570\,,
$$
where $A$ is the mass number of the isotope.
We ascribed the uncertainty of 1\% to this approximate formula, which was intended as an order-of-magnitude estimate.
An update of the charge radii of light nuclei with $Z$ up to 32 was recently published by Ohayon \cite{ohayon:25:radii},
see also Ref.~\cite{beyer:25}.
We do not incorporate this update in the present work,
as our transition energies are not yet
sensitive enough to nuclear radii for these light elements.

It is important to note, however, that Ref.~\cite{ohayon:25:radii} raised concerns about the reliability of the uncertainty estimates in the compilation by Angeli and Marinova.
In particular, Ohayon argued that the model dependence of the nuclear charge distribution was
not properly accounted for,
suggesting that the uncertainties reported in that work should be increased by a factor of 2 to 3.
A similar conclusion was reached in a recent reevaluation of the nuclear charge radius of $^{238}$Pb
based on old muonic spectroscopy data and an updated QED theory \cite{sun:25}.
In the present work, we continue to use the nuclear charge radii by Angeli and Marinova,
while bearing in mind that the associated uncertainties may be underestimated.

Following Ref.~\cite{yerokhin:15:Hlike}, the uncertainties of the fns effect for all ions except uranium
were evaluated
by quadratically combining two contributions:
(i) the uncertainty due to variation of the nuclear radii within their error bars and
(ii) the difference between the fns results obtained with the Fermi model and the uniformly charged sphere model for the nuclear charge distribution.
The comparison between the Fermi and uniform models yields probably an overly
conservative estimate of the model dependence for the spherical nuclei;
however, it is supposed also to cover possible nuclear deformation effects for
deformed nuclei.
This estimate
can be improved by a careful examination of the nuclear charge model
and dedicated calculations for a particular isotope \cite{valuev:20}.

An example of a detailed examination  of the fns effect
was performed for the experimentally important case
of $^{238}$U in the
work by Kozhedub et al.~\cite{kozhedub:08}. It was shown that for this isotope
the quadrupole and hexadecapole deformations lead to sizeable nuclear-deformation contributions
to the fns effect. As a result, Kozhedub et al.~arrived at the fns uncertainty of 0.030~eV for
the $2p_{1/2}$--$2s$ transition energy, which may be compared to the uncertainty of 0.076~eV delivered
by the standard estimate.
In this work we adopt the results for the nuclear-deformation corrections and
the uncertainties of the fns effect for $^{238}$U from Kozhedub et al.

The nuclear-polarization correction was calculated for selected isotopes in
Refs.~\cite{plunien:91,plunien:95,nefiodov:96,nefiodov:02:prl}.
Following Ref.~\cite{yerokhin:15:Hlike}, we use the calculated results when available,
with the ascribed uncertainty of 50\%, and a crude estimate from
Ref.~\cite{yerokhin:15:Hlike} with the uncertainty of 100\% in all other cases.

\section{Results and discussion}
\label{sec:results}

Individual theoretical contributions to the transition energies of Li-like ions are summarized in
Table~\ref{tab:breakdwn}.
The second and third columns of the table list our numerical values for the DCB energy
and the total electron-structure (STRUC) energy which includes both the DCB and QED electron-structure contributions.
We note  that the DCB energies slightly depend on the choice of the screening potential in the
one-electron Dirac Hamiltonian; the results presented were obtained using the LDF potential.
In contrast, the electron-structure energies should not depend on the starting potential
within the given uncertainties.

The fourth column (QED1) of the table lists results for the one-electron one-loop QED corrections; they were
taken from the compilation \cite{yerokhin:15:Hlike}. The next column (QED1SCR)
contains results for the self-energy and vacuum-polarization screening corrections
evaluated in Sec.~\ref{sec:qedscr}. The QED2 column shows results for the two-loop QED correction,
including the estimated screening effect, see  Sec.~\ref{sec:qedscr} for details.
The REC column contains results for the nuclear recoil correction obtained in Sec.~\ref{sec:rec}.
The NP column gives the nuclear polarization correction obtained
as described in Sec.~\ref{sec:nucl}. For $Z = 92$, the NP entry additionally includes
the nuclear-deformation correction
of $-0.026$~eV and $-0.003$~eV for the $2s$, and
$2p_{1/2}$ states, correspondingly \cite{kozhedub:08}.

We observe that the source of the dominant theoretical uncertainty varies with the nuclear charge $Z$.
For $Z < 20$, the theoretical uncertainty is primarily due to the DCB energy.
In the range $20 < Z < 40$, it is mainly determined by the QED screening and QED electron-structure effects.
For $Z>40$, however, the largest theoretical error is coming from the one-electron two-loop QED effects.
We conclude that once the complete evaluation of all two-loop one-electron QED corrections to all orders in $\Za$
is finished \cite{volkov:25:tobe}, it will improve the theoretical
precision not only for hydrogen-like ions, but also for Li-like ions with $Z>40$.

Table~\ref{tab:total} summarizes our final theoretical results for the ionization energies
of the $2s$, $2p_{1/2}$, and $2p_{3/2}$ states and
the $2p_{1/2}$--$2s$ and $2p_{3/2}$--$2s$ transition energies of Li-like ions with
nuclear charges $Z = 10$--$100$.
The theoretical values are specified with typically two uncertainties, where the first one
is the purely theoretical one and the second
represents the estimated error due to the nuclear charge distribution.
As seen from the table, the second uncertainty becomes significant for $Z > 60$.
Note that for $Z \le 20$, the uncertainties of ionization energies are larger than those
of transition energies. This is because the DCB transition
energies for these ions
were obtained by the CI method, whereas for ionization energies the DCB part was calculated with MBPT.

For many nuclei in the high-$Z$ region,
the nuclear charge radii are not very well known, with typical errors of about 1\% or worse.
In such cases, the uncertainty due to the nuclear radius dominates the total theoretical error,
indicating that the accuracy of these radii can be improved by measuring the transition
energies in the corresponding Li-like ions and comparing with our theoretical
predictions.

For other isotopes, the nuclear radii are known with higher precision
reaching $0.1\%$ or even better \cite{angeli:13}.
For such ions, the uncertainty due to the nuclear radius
in theoretical transition energies is comparable to, but still
smaller than the purely theoretical uncertainty.
This means that our calculations cannot yet provide a determination of
nuclear radii with the precision that rivals the accuracy claimed in Ref.~\cite{angeli:13}.
However, such determination will become possible in the future,
once the project of the complete
evaluation of all one-electron two-loop QED effects is completed and
the dominant source of theoretical uncertainty is eliminated.

We now turn to the comparison of our theoretical results for
the $2p_{1/2}$--$2s$ and $2p_{3/2}$--$2s$ transition energies
with previous theoretical calculations and available experimental data.
Such comparison is presented in  Table~\ref{tab:comparison}.
We find excellent agreement with the calculation of Kozhedub et
al.~\cite{kozhedub:10}. The differences between the calculated values
are always within the previously estimated
error bars, while our results are
more accurate. There are, however,
small but noticeable deviations from theoretical values by
Sapirstein and Cheng \cite{sapirstein:11}.

Table~\ref{tab:comparison} also provides a comparison between theoretical predictions
and available experimental data, demonstrating generally very good agreement.
We conclude that our current theoretical predictions surpass in accuracy nearly
all existing experimental results for $Z\ge 10$, with only a few exceptions.
One exception is the measurement of the $2p_{3/2}$--$2s$ transition energy
in neon by Bockasten et al.~from 1963  \cite{bockasten:63}. Already at that time
the experimental
accuracy of 0.10~meV has been achieved, which is slightly better than our theoretical uncertainty of 0.16~meV,
and their result is in excellent agreement with our theoretical value.

The second instance where the experimental precision surpasses that of theory
lies at the opposite end of Periodic Table.
Namely, the
$2p_{1/2}$--$2s$ transition energy in uranium was measured by Beiersdorfer et
al.~\cite{beiersdorfer:05} with an outstanding accuracy of 15~meV, which
might be compared to our purely theoretical uncertainty of 71~meV.
We observe a small tension between theory and experiment in this case,
with a deviation of about 1.5 times the theoretical uncertainty.
This might be an indication that the nuclear charge radius of $^{238}$U
is known to a lesser extent than was believed in Ref.~\cite{angeli:13}.
Indeed, if the uncertainty of the
nuclear radius is increased by a factor of 3, as suggested in Ref.~\cite{ohayon:25:radii},
the difference between theory and experiment would be reduced to 1$\sigma$.

Another important experimental result is the measurement of the
$2p_{3/2}$--$2s$ transition energy in $^{209}$Bi by
Beiersdorfer et al.~\cite{beiersdorfer:98}.
The experimental accuracy of 40~meV matches that of
our current theoretical precision, and excellent agreement is observed.
Further measurements in the high‑$Z$ range are listed in the table,
though their larger uncertainties make them less precise than our theoretical predictions.

\section{Comparison with approximate QED treatment}
\label{sec:qedmod}

For atomic systems with many electrons, an {\em ab initio} treatment of QED
effects is generally not feasible. Consequently,
approximate methods based on various
versions of so-called QED potentials are typically used for such systems.
Among these, the most successful approach is based on
the model QED operator introduced by Shabaev et al.~\cite{shabaev:13:qedmod}
and implemented in the QEDMOD package \cite{shabaev:14:qedmod}.
The accuracy of the model QED operator method  can only be assessed by
comparison with {\em ab initio} QED calculations. So,
we now perform an analysis of the model QED operator’s performance, aiming to
provide a basis for estimating its accuracy for many-electron calculations.

We computed approximate QED contributions
to transition energies of Li-like ions by adding
the QEDMOD operator from Ref.~\cite{shabaev:14:qedmod}
to the DCB Hamiltonian in our CI calculations
and evaluating the difference between results obtained with and without the
QEDMOD addition. Following Ref.~\cite{yerokhin:14:belike},
we include matrix elements of the QEDMOD operator only between one-electron
states that lie below the continuum energy threshold.

Table~\ref{tab:qedmod} presents our results
obtained with QEDMOD operator ($E_{\rm mqed}$) in comparison
with the {\em ab initio} QED corrections evaluated in the preceding sections.
$E_{\rm mqed}$ is intended to approximately reproduce the
radiative QED correction $E_{\rm qed,rad}$.
For comparison, the
table also includes the electron-structure QED contributions
not accounted for by the QEDMOD operator, namely, the one-photon
and two-photon exchange corrections, $E^{(1)}_{\rm qed}$ and
$E^{(2)}_{\rm qed}$. We note that $E^{(1)}_{\rm qed}$ can be
incorporated in many-body calculations via
the so-called frequency-dependent Breit correction, see, e.g.,
Ref.~\cite{yerokhin:18:lilike} for details.

We observe that for the $2p_{1/2}$--$2s$ and $2p_{3/2}$--$2s$ transition
energies, the QEDMOD operator reproduces the {\em ab initio} values of the
radiative QED correction $E_{\rm qed,rad}$ very well, with differences
within 2\% in nearly all cases.
This accuracy is remarkable, having in mind that $E_{\rm qed,rad}$ includes
also two-loop QED effects, which are not accounted for by the QEDMOD operator.
The one-photon electron-structure QED correction $E^{(1)}_{\rm qed}$
contributes at the level of just 1\% for the $2p_{1/2}$--$2s$ transition,
but increases to as much as
20\% for the $2p_{3/2}$--$2s$ transition.
We therefore conclude that this correction should be included alongside
the QEDMOD operator to achieve an adequate representation of QED effects
in many-body calculations.
In contrast,  the two-photon QED contribution introduces a small correction
on the level of 0.5\% for $2p$--$2s$ transitions and its omission does not
compromise the accuracy of the QEDMOD operator.

The situation for the fine-structure $2p_{3/2}$--$2p_{1/2}$ transition
is markedly  different.
Radiative QED effects are strongly suppressed in this fine-structure difference,
and as a result, the {\em relative} accuracy of the QEDMOD approximation is significantly
lower. For most ions, the deviation remains within 10\%, but it
worsens in the low-$Z$ region, reaching up to 30\% for $Z = 10$.
It is worth noting that performance of the QEDMOD approach in this regime can
be substantially improved by supplementing it with anomalous magnetic moment (AMM)
operators, resulting in the combined QEDMOD$+$AMM treatment~\cite{yerokhin:20:fs}.

The one-photon electron-structure QED correction $E^{(1)}_{\rm qed}$ is also greatly
enhanced for the fine-structure transition and even becomes the dominant QED contribution
in the high-$Z$ region. The two-photon correction $E^{(2)}_{\rm qed}$
ranges between 2\% and 9\%, its magnitude being comparable to the deviation between the
QEDMOD values and the exact radiative corrections.
We conclude that, with the frequency-dependent Breit correction taken into account,
the accuracy of the QEDMOD treatment for the fine-structure
transition remains within 10\% for nuclear charges $Z\ge 30$, but gradually declines
for lower values of $Z$.

\section*{Summary}

We have performed systematic QED calculations of the ionization energies of the
$2s$, $2p_{1/2}$, and $2p_{3/2}$ states and the $2p_{1/2}$--$2s$ and $2p_{3/2}$--$2s$
transition energies for Li-like ions with the nuclear charge numbers $Z = 10$--$100$.
{\em Ab initio} QED calculations were carried out for the QED screening effects,
the QED electron-structure effects with one and two photon exchanges,
and the nuclear recoil effect.
In order to improve convergence of our QED calculations, we employed the
extended Furry picture, starting with the Dirac equation with a local
screening potential.
Higher-order electron-structure effects were accounted for within
the Breit approximation, by solving the no-pair Dirac-Coulomb-Breit Hamiltonian
with the MBPT and CI methods.
The QED screening effects with two photon exchanges
were approximately accounted for by using the model QED operator.

The obtained theoretical predictions improve upon the best previous
QED calculations of the $2p_{1/2}$--$2s$ and $2p_{3/2}$--$2s$
transition energies \cite{kozhedub:10,sapirstein:11}.
It has been demonstrated that the present
theoretical energies surpass in accuracy nearly
all existing experimental results for $Z\ge 10$,
with exception of the $Z = 10$ measurement of Ref.~\cite{bockasten:63}
and the $Z = 92$ result of Ref.~\cite{beiersdorfer:05}.

Comparison of our theoretical predictions
for uranium and bismuth
with the available experimental data
yields one of the best tests of bound-state QED theory in the region
of a strong nuclear binding field.
Alternatively, this comparison can be used for an accurate determination of the
nuclear charge radii.
The current theoretical precision is sufficient to determine nuclear
radii of high-$Z$ ions with accuracy on the level of 1\%, which
is of importance for many nuclei with not-too-well studied charge distribution.
In order to reach the precision on the level of $0.1$\% in nuclear radii
determinations, claimed in Ref.~\cite{angeli:13} for the
best studied isotopes, one needs to complete the calculations
of all one-electron two-loop QED effects, which will improve
substantially the theoretical accuracy.

\section*{Acknowledgement}

Valuable discussions with J.~R.~Crespo L\'opez-Urrutia are
gratefully acknowledged.

%
%
%
\begin{table*}[htb]
\begin{center}
\caption{
Breakdown of theoretical calculations of the $2p_{1/2}$--$2s$ and $2p_{3/2}$--$2s$ transition energies of Li-like ions, in eV.
Abbreviations are as follows: DCB, the Dirac-Coulomb-Breit energy; STRUC, the electron-structure energy
(consisting of the DCB and QED electron-structure parts);
QED1, the one-electron one-loop QED contribution; QED1SCR, the one-loop QED screening correction;
QED2, the two-loop QED correction (including the estimate of screening); REC, the nuclear recoil correction; NP, the nuclear polarization;
Total, the total theory. Uncertainties due to the nuclear radii are shown only for the total theory values.
There, the first uncertainty is the purely theoretical error; the second uncertainty (if present) is the error due to
the nuclear charge distribution.
\label{tab:breakdwn}
}
\begin{ruledtabular}
\begin{tabular}{lw{4.8}w{2.8}w{2.8}w{1.8}w{1.8}w{2.8}w{2.8}w{2.8}w{2.11}}
$Z$
   & \multicolumn{1}{c}{DCB} & \multicolumn{1}{c}{STRUC}
                               & \multicolumn{1}{c}{QED1}
                                          & \multicolumn{1}{c}{QED1SCR}
                                           & \multicolumn{1}{c}{QED2}
                                           & \multicolumn{1}{c}{REC}
                                           & \multicolumn{1}{c}{NP}
                                           & \multicolumn{1}{c}{Total}
\\\hline\\[-9pt]
  \multicolumn{5}{l}{$2p_{1/2}$--$2s$} \\
 10    &  15.9070\,(2)      &  15.9069\,(2)      & -0.02002           &  0.00585\,(6)      &                    & -0.00423           &                    &  15.88854\,(17)    \\
 12    &  19.8730\,(2)      &  19.8728\,(2)      & -0.03811           &  0.00934\,(8)      &  0.00001           & -0.00539           &                    &  19.83865\,(19)    \\
 14    &  23.8709\,(2)      &  23.8706\,(2)      & -0.06545           &  0.0138\,(1)       &  0.00003\,(1)      & -0.00655\,(1)      &                    &  23.81242\,(22)    \\
 18    &  32.0069\,(2)      &  32.0062\,(2)      & -0.15698           &  0.0260\,(1)       &  0.00009\,(1)      & -0.00799\,(1)      &                    &  31.86732\,(29)    \\
 20    &  36.1645\,(2)      &  36.1635\,(3)      & -0.22597           &  0.0338\,(2)       &  0.00015\,(2)      & -0.01004\,(2)      &                    &  35.96140\,(33)    \\
 22    &  40.3943\,(2)      &  40.3927\,(3)      & -0.31380           &  0.0428\,(2)       &  0.00025\,(3)      & -0.01027\,(2)      &                    &  40.11162\,(38)    \\
 26    &  49.1067\,(1)      &  49.1034\,(3)      & -0.55650           &  0.0646\,(2)       &  0.00054\,(7)      & -0.01258\,(4)      &  0.00001\,(1)      &  48.59947\,(37)(1) \\
 30    &  58.2196\,(1)      &  58.2135\,(3)      & -0.90697           &  0.0916\,(3)       &  0.0011\,(1)       & -0.01491\,(6)      &  0.00001\,(1)      &  57.38430\,(43)(1) \\
 36    &  72.8155\,(1)      &  72.8018\,(3)      & -1.68595           &  0.1426\,(4)       &  0.0025\,(3)       & -0.01672\,(9)      &  0.00002\,(1)      &  71.24426\,(60)(5) \\
 40    &  83.2928\,(1)      &  83.2708\,(4)      & -2.41068           &  0.1839\,(5)       &  0.0041\,(5)       & -0.0195\,(1)       &  0.00007\,(7)      &  81.02860\,(78)(7) \\
 50    &  112.8043\,(1)     &  112.7447\,(6)     & -5.14309           &  0.3141\,(8)       &  0.011\,(2)        & -0.0238\,(2)       &  0.0003\,(3)       &  107.9037\,(19)(4) \\
 54    &  126.2295          &  126.1461\,(7)     & -6.6851            &  0.377\,(1)        &  0.016\,(2)        & -0.0257\,(3)       &  0.0004\,(5)       &  119.8295\,(27)(10) \\
 60    &  148.5109          &  148.381\,(1)      & -9.5874            &  0.484\,(1)        &  0.027\,(4)        & -0.0304\,(5)       &  0.0004\,(2)       &  139.2746\,(46)(15) \\
 70    &  192.343           &  192.107\,(2)      & -16.330            &  0.693\,(3)        &  0.06\,(1)         & -0.036\,(1)        &  0.003\,(2)        &  176.492\,(11)(7) \\
 83    &  264.807           &  264.433\,(3)      & -29.723\,(1)       &  1.000\,(7)        &  0.13\,(3)         & -0.050\,(2)        &  0.01\,(1)         &  235.802\,(36)(26) \\
 90    &  309.663           &  309.287\,(4)      & -39.680\,(2)       &  1.16\,(1)         &  0.20\,(6)         & -0.059\,(2)        &  0.02\,(1)         &  270.921\,(60)(106) \\
 92    &  322.638           &  322.286\,(5)      & -42.929\,(1)       &  1.19\,(2)         &  0.22\,(7)         & -0.063\,(3)        &  0.06\,(2)^a         &  280.767\,(72)(34) \\
  \multicolumn{5}{l}{$2p_{3/2}$--$2s$} \\
%
 10    &  16.1112\,(2)      &  16.1111\,(2)      & -0.01899           &  0.00531\,(6)      &                    & -0.00424           &                    &  16.09316\,(17)    \\
 12    &  20.3653\,(2)      &  20.3650\,(2)      & -0.03598           &  0.00839\,(8)      &  0.00001           & -0.00541           &                    &  20.33200\,(19)    \\
 14    &  24.8828\,(2)      &  24.8821\,(2)      & -0.06153           &  0.0123\,(1)       &  0.00002\,(1)      & -0.00658\,(1)      &                    &  24.82636\,(21)    \\
 18    &  35.1714\,(2)      &  35.1694\,(2)      & -0.14641           &  0.0228\,(2)       &  0.00007\,(1)      & -0.00805\,(1)      &                    &  35.03775\,(28)    \\
 20    &  41.2192\,(2)      &  41.2157\,(3)      & -0.20999           &  0.0294\,(2)       &  0.00013\,(2)      & -0.01013\,(2)      &                    &  41.02510\,(32)    \\
 22    &  48.0856\,(2)      &  48.0800\,(2)      & -0.29058           &  0.0370\,(2)       &  0.00020\,(3)      & -0.01039\,(2)      &                    &  47.81627\,(33)    \\
 26    &  65.0466\,(1)      &  65.0339\,(2)      & -0.51193           &  0.0553\,(3)       &  0.00046\,(7)      & -0.01279\,(4)      &  0.00001\,(1)      &  64.56497\,(35)(1) \\
 30    &  87.8192\,(1)      &  87.7933\,(2)      & -0.82938           &  0.0780\,(3)       &  0.0009\,(1)       & -0.01525\,(6)      &  0.00001\,(1)      &  87.02766\,(42)(1) \\
 36    &  137.6694\,(1)     &  137.6050\,(3)     & -1.52983           &  0.1209\,(5)       &  0.0021\,(3)       & -0.01730\,(9)      &  0.00002\,(1)      &  136.18092\,(64)(5) \\
 40    &  185.2575\,(1)     &  185.1481\,(4)     & -2.17805           &  0.1561\,(5)       &  0.0035\,(5)       & -0.0204\,(1)       &  0.00007\,(7)      &  183.10933\,(84)(7) \\
 50    &  379.3688\,(1)     &  379.0333\,(7)     & -4.61298           &  0.2713\,(8)       &  0.010\,(2)        & -0.0255\,(2)       &  0.0003\,(3)       &  374.6762\,(19)(4) \\
 54    &  498.3894\,(1)     &  497.8955\,(9)     & -5.9882            &  0.3301\,(9)       &  0.014\,(2)        & -0.0279\,(3)       &  0.0005\,(5)       &  492.2242\,(27)(10) \\
 60    &  738.2029          &  737.365\,(1)      & -8.5886            &  0.436\,(1)        &  0.023\,(4)        & -0.0338\,(5)       &  0.0004\,(2)       &  729.2020\,(46)(15) \\
 70    &  1361.383          &  1359.567\,(2)     & -14.727            &  0.672\,(2)        &  0.05\,(1)         & -0.042\,(1)        &  0.003\,(2)        &  1345.521\,(11)(7) \\
 83    &  2818.645          &  2814.395\,(3)     & -27.486\,(1)       &  1.140\,(3)        &  0.12\,(3)         & -0.062\,(2)        &  0.01\,(1)         &  2788.116\,(35)(26) \\
 90    &  4067.713          &  4061.360\,(4)     & -37.572\,(2)       &  1.505\,(5)        &  0.18\,(6)         & -0.077\,(2)        &  0.02\,(1)         &  4025.421\,(58)(105) \\
 92    &  4505.820          &  4498.738\,(4)     & -40.991\,(1)       &  1.628\,(6)        &  0.21\,(7)         & -0.083\,(3)        &  0.07\,(2)^a         &  4459.565\,(69)(34) \\
\end{tabular}
\end{ruledtabular}
\end{center}
$^a$ includes the nuclear-deformation correction from Ref.~\cite{kozhedub:08}.
\end{table*}

%
%
%
\begin{table*}[htb]
\begin{center}
\caption{
Comparison of theoretical and experimental results for transition energies of Li-like ions, in eV.
\label{tab:comparison}
}
\begin{ruledtabular}
\begin{tabular}{lw{3.12}w{3.10}w{3.10}cw{3.12}w{3.10}w{3.10}c}

              &  \multicolumn{4}{c}{$2p_{1/2}$--$2s$}
                                                        &  \multicolumn{4}{c}{$2p_{3/2}$--$2s$}
\\
\cline{2-5}\cline{6-9}
\\[-5pt]
$Z$
   & \multicolumn{1}{c}{This work} & \multicolumn{1}{c}{Other theory} & \multicolumn{2}{c}{Experiment}
   & \multicolumn{1}{c}{This work} & \multicolumn{1}{c}{Other theory} & \multicolumn{2}{c}{Experiment}
\\\hline\\[-9pt]
 10 &    15.88854\,(17)     & 15.8883\,(4)^a & 15.8888\,(2)   & \cite{bockasten:63}     &  16.09316\,(17)     & 16.0932\,(4)^a   & 16.09330\,(10) & \cite{bockasten:63} \\ 
    &                       & 15.8881\,(5)^b & 15.8887\,(3)   & \cite{edlen:83}         &                     & 16.0923\,(5)^b   & 16.09315\,(35) &  \cite{edlen:83} \\
 11 &    17.86153\,(18)     &                & 17.8614\,(4)   & \cite{edlen:83}         &  18.18708\,(18)     &                  & 18.1876\,(5)   &  \cite{edlen:83} \\
 12 &    19.83865\,(19)     &                & 19.8390\,(4)   & \cite{edlen:83}         &  20.33200\,(19)     &                  & 20.3318\,(5)   &  \cite{edlen:83} \\
 13 &    21.82216\,(20)     &                & 21.8227\,(5)   & \cite{edlen:83}         &  22.54100\,(20)     &                  & 22.5413\,(7)   &  \cite{edlen:83} \\
 14 &    23.81242\,(22)     &                & 23.8125\,(4)   & \cite{edlen:83}         &  24.82636\,(21)     &                  & 24.8264\,(5)   &  \cite{edlen:83} \\
 15 &    25.81139\,(23)     & 25.8110\,(4)^a & 25.8098\,(15)  & \cite{edlen:83}         &  27.20316\,(23)     & 27.2026\,(5)^a   & 27.205\,(2)   &  \cite{edlen:83} \\
 16 &    27.81913\,(17)     &                & 27.8187\,(7)   & \cite{edlen:83}         &  29.68574\,(25)     &                  & 29.6863\,(11) &  \cite{edlen:83} \\
 18 &    31.86732\,(29)     & 31.8673\,(5)^a & 31.8664\,(9)   & \cite{edlen:83}         &  35.03775\,(28)     & 35.0378\,(6)^a   & 35.0380\,(6)  & \cite{peacock:84}\\
 20 &    35.96140\,(33)     & 35.9612\,(6)^a & 35.9614\,(10)  & \cite{edlen:83}         &  41.02510\,(32)     & 41.0251\,(7)^a   & 41.0261\,(14) &  \cite{edlen:83} \\
    &                       & 35.962\,(1)^c  &                &                         &                     & 41.024\,(1)^c \\
 21 &    38.02958\,(39)     & 38.0289\,(7)^a &                &                         &  44.30964\,(33)     & 44.3092\,(7)^a   & 44.3094\,(2)  & \cite{lestinsky:08}\\ 
    &                       & 38.031\,(1)^c  &                &                         &                     & 44.308\,(1)^c\\
 22 &    40.11162\,(38)     &                & 40.1150\,(12)  & \cite{peacock:84}       &  47.81627\,(33)     &                  & 47.8201\,(7)  & \cite{peacock:84}\\
 24 &    44.32185\,(37)(1)  &                & 44.328\,(4)    & \cite{edlen:83}         &  55.59376\,(33)(1)  &                  & 55.5936\,(15) & \cite{knize:91}\\
    &                       &                & 44.323\,(3)    & \cite{hinnov:89}        & & & & \\
 25 &    46.45198\,(37)(1)  &                & 46.459\,(5)    & \cite{edlen:83}         &  59.91654\,(34)(1)  &                  & 59.928\,(7)   &  \cite{edlen:83} \\
 26 &    48.59947\,(37)(1)  & 48.5991\,(9)^a & 48.5982\,(8)   & \cite{epp:07}           &  64.56498\,(35)(1)  & 64.5650\,(9)^a   & 64.566\,(2)   & \cite{reader:94}  \\
    &                       & 48.599\,(1)^c  & 48.5997\,(10)  & \cite{reader:94}        &                     & 64.562\,(1)^c    & 64.560\,(3)   & \cite{knize:91}\\
 28 &    52.95113\,(38)(1)  & 52.9504\,(10)^a& 52.9501\,(11)  & \cite{sugar:92,sugar:93}&  74.95868\,(38)(1)  & 74.9586\,(11)^a  & 74.960\,(2)   & \cite{sugar:92,sugar:93} \\
    &                       & 52.951\,(1)^c  & 52.9496\,(23)  & \cite{hinnov:89}        &                     & 74.955\,(1)^c    & 74.962\,(5)   & \cite{hinnov:89} \\
 30 &    57.38430\,(43)(1)  & 57.3846\,(10)^a& 57.384\,(3)  & \cite{staude:98}          &  87.02766\,(42)(1)  & 87.0282\,(12)^a  & 87.030\,(4)   & \cite{staude:98} \\
    &                       & 57.382\,(1)^c   &                &                        &                     & 87.023\,(1)^c  \\
 32 &    61.90632\,(54)(2)  & 61.904\,(1)^c  & 61.901\,(2)  & \cite{knize:91}           &  101.04873\,(49)(2) &                  & 101.043\,(5)  & \cite{knize:91}\\
 36 &    71.24426\,(60)(5)  & 71.2451\,(15)^a& 71.243\,(8)    & \cite{madzunkov:02}     &  136.18092\,(64)(5) & 136.1818\,(17)^a & 136.16\,(3) & \cite{denne:87} \\
    &                       & 71.240\,(1)^c  &               &                          &                     &                  & 136.17\,(4) & \cite{hinnov:89} \\
 39 &    78.53716\,(71)(7)  &                & 78.540\,(5)    & \cite{silwal:17}        &  170.10314\,(79)(7) &                  & 170.135\,(14) & \cite{silwal:17} \\
 42 &    86.1072\,(13)(1)   & 86.104\,(2)^c  & 86.101\,(12)   & \cite{hinnov:89}        &  211.9868\,(10)(1)  &                  & 211.94\,(7) & \cite{hinnov:89} \\
 47 &    99.4318\,(16)(3)   & 99.432\,(4)^a  & 99.438\,(7)   & \cite{bosselmann:99}      &  303.6707\,(15)(3)  & 303.6709\,(36)^a & 303.67\,(3) & \cite{bosselmann:99} \\
    &                       & 99.414\,(3)^c  &                &                         &                     &  \\
 50 &    107.9037\,(19)(4)  & 107.904\,(5)^a & 107.911\,(8)  & \cite{feili:00}          &  374.6762\,(19)(4)  & 492.225\,(6)^a   & & \\
 54 &    119.8295\,(27)(10) & 119.831\,(6)^a & 119.820\,(8)  & \cite{feili:00}          &  492.2242\,(27)(10) &                  & 492.17\,(5) & \cite{bernhardt:15} \\
 56 &    126.0909\,(33)(12) &                & 126.112\,(13) &  \cite{reader:14}        &  562.3505\,(33)(12) &                  & & \\
 74 &    193.410\,(17)(9)   & 193.44\,(3)^c  &                &                         &  1696.121\,(16)(9)  & 1696.10\,(3)^c   & 1696.2\,(5) & \cite{clementson:11} \\
 79 &    216.277\,(25)(17)  & 216.22\,(3)^c  & 216.13\,(10)   & \cite{brandau:03}       &  2244.027\,(25)(17) & 2244.00\,(3)^c   & & \\
 82 &    230.817\,(31)(22)  & 230.76\,(4)^c  & 230.65\,(8)    & \cite{brandau:03}       &  2642.210\,(30)(22) & 2642.17\,(4)^c   &  2642.26\,(10) & \cite{zhang:08} \\
 83 &    235.802\,(36)(26)  & 235.72\,(5)^c  &                &                         &  2788.116\,(35)(26) & 2788.04\,(5)^c   & 2788.14\,(4) & \cite{beiersdorfer:98} \\
    &                       &                &                &                         &                     & 2788.12\,(7)^d \\
 90 &    270.921\,(60)(106) & 270.74\,(7)^c  &                &                         &  4025.421\,(58)(105)& 4025.25\,(7)^c   & 4025.23\,(15) & \cite{beiersdorfer:95} \\
 92 &    280.767\,(72)(34)  & 280.65\,(8)^c  & 280.645\,(15)  & \cite{beiersdorfer:05}  &  4459.565\,(69)(34) & 4459.57\,(10)^f & 4459.4\,(2) & \cite{beiersdorfer:93}\\
    &                       & 280.76\,(14)^d &                &                         &                     & 4459.46\,(8)^c \\
\end{tabular}
\end{ruledtabular}
\end{center}
$^a$ Kozhedub et al. (2010) \cite{kozhedub:10}, \\
$^b$ Wang et al. (2023) \cite{wang:23},\\
$^c$ Sapistein and Cheng (2011) \cite{sapirstein:11},\\
$^d$ Yerokhin et al. (2006) \cite{yerokhin:06:prl}, \\
$^f$ Malyshev, Kozhedub, and Shabaev (2023) \cite{malyshev:23}. \\
\end{table*}

%
%
%
\begin{longtable*}{llllw{6.12}w{6.12}w{6.12}w{3.12}w{4.12}}
\caption{
Theoretical ionization and transition energies of Li-like ions, in eV.
When two uncertainties are specified, the first represents the estimated theoretical
uncertainty, and the second is the uncertainty induced by the nuclear charge distribution. If only a
single uncertainty is given, the error from the nuclear size is negligible.
$R$ is the nuclear charge radius used in calculations.
\label{tab:total}
}
\\ \colrule\hline
$Z$ & $A$ &
& \multicolumn{1}{c}{$R$ [fm]}
   & \multicolumn{1}{c}{$2s$} & \multicolumn{1}{c}{$2p_{1/2}$} & \multicolumn{1}{c}{$2p_{3/2}$}
    & \multicolumn{1}{c}{$2p_{1/2}$--$2s$}
    & \multicolumn{1}{c}{$2p_{3/2}$--$2s$}
\\\hline\\[-9pt]
%
 10 &  20  & Ne   & 3.006\,(2)  & -239.0979\,(18)        & -223.2089\,(27)        &  -223.0043\,(26)       &  15.88854\,(17)        &  16.09316\,(17)    \\
 11 &  23  & Na   & 2.994\,(2)  & -299.8657\,(16)        & -282.0038\,(23)        &  -281.6783\,(22)       &  17.86153\,(18)        &  18.18708\,(18)    \\
 12 &  24  & Mg   & 3.057\,(2)  & -367.5003\,(13)        & -347.6614\,(19)        &  -347.1680\,(18)       &  19.83865\,(19)        &  20.33200\,(19)    \\
 13 &  27  & Al   & 3.061\,(3)  & -442.0165\,(11)        & -420.1942\,(17)        &  -419.4753\,(16)       &  21.82216\,(20)        &  22.54100\,(20)    \\
 14 &  28  & Si   & 3.122\,(2)  & -523.42793\,(97)       & -499.6154\,(15)        &  -498.6014\,(14)       &  23.81242\,(22)        &  24.82636\,(21)    \\
 15 &  31  & P    & 3.189\,(2)  & -611.75244\,(87)       & -585.9409\,(13)        &  -584.5491\,(12)       &  25.81139\,(23)        &  27.20316\,(23)    \\
 16 &  32  & S    & 3.261\,(2)  & -707.00662\,(77)       & -679.1874\,(12)        &  -677.3207\,(11)       &  27.81913\,(17)        &  29.68574\,(25)    \\
 17 &  35  & Cl   & 3.37\,(2)   & -809.21143\,(71)(1)    & -779.3738\,(11)        &  -776.91988\,(96)      &  29.83748\,(27)(1)     &  32.29142\,(26)(1) \\
 18 &  40  & Ar   & 3.427\,(3)  & -918.38787\,(64)       & -886.52045\,(97)       &  -883.35000\,(87)      &  31.86732\,(29)        &  35.03775\,(28)    \\
 19 &  39  & K    & 3.435\,(2)  & -1034.55575\,(60)      & -1000.64786\,(90)      &  -996.61387\,(80)      &  33.90779\,(31)        &  37.94176\,(30)    \\
 20 &  40  & Ca   & 3.478\,(2)  & -1157.74186\,(56)      & -1121.78036\,(83)      &  -1116.71663\,(73)     &  35.96140\,(33)        &  41.02510\,(32)    \\
 21 &  45  & Sc   & 3.546\,(3)  & -1287.97265\,(54)      & -1249.94307\,(80)      &  -1243.66301\,(70)     &  38.02958\,(39)        &  44.30964\,(33)    \\
 22 &  48  & Ti   & 3.592\,(2)  & -1425.27317\,(52)      & -1385.16155\,(76)      &  -1377.45689\,(65)     &  40.11162\,(38)        &  47.81627\,(33)    \\
 23 &  51  & V    & 3.600\,(2)  & -1569.67304\,(49)      & -1527.46410\,(72)      &  -1518.10360\,(61)     &  42.20895\,(37)        &  51.56945\,(32)    \\
 24 &  52  & Cr   & 3.645\,(4)  & -1721.20193\,(47)(1)   & -1676.88008\,(68)      &  -1665.60817\,(56)     &  44.32185\,(37)(1)     &  55.59376\,(33)(1) \\
 25 &  55  & Mn   & 3.706\,(2)  & -1879.89324\,(46)(1)   & -1833.44126\,(66)      &  -1819.97669\,(54)     &  46.45198\,(37)(1)     &  59.91654\,(34)(1) \\
 26 &  56  & Fe   & 3.738\,(2)  & -2045.77956\,(46)(1)   & -1997.18009\,(64)      &  -1981.21459\,(52)     &  48.59947\,(37)(1)     &  64.56497\,(35)(1) \\
 27 &  59  & Co   & 3.788\,(2)  & -2218.89769\,(45)(1)   & -2168.13177\,(62)      &  -2149.32857\,(50)     &  50.76592\,(37)(1)     &  69.56912\,(36)(1) \\
 28 &  58  & Ni   & 3.776\,(2)  & -2399.28294\,(45)(1)   & -2346.33182\,(61)      &  -2324.32426\,(49)     &  52.95113\,(38)(1)     &  74.95868\,(38)(1) \\
 29 &  63  & Cu   & 3.882\,(2)  & -2586.97731\,(46)(1)   & -2531.81985\,(60)      &  -2506.20987\,(49)     &  55.15746\,(40)(1)     &  80.76745\,(40)(1) \\
 30 &  64  & Zn   & 3.928\,(2)  & -2782.01868\,(47)(1)   & -2724.63438\,(61)      &  -2694.99102\,(49)     &  57.38430\,(43)(1)     &  87.02766\,(42)(1) \\
 31 &  69  & Ga   & 3.997\,(2)  & -2984.45226\,(49)(2)   & -2924.81835\,(68)      &  -2890.67619\,(48)     &  59.63391\,(49)(2)     &  93.77607\,(45)(2) \\
 32 &  74  & Ge   & 4.074\,(1)  & -3194.32119\,(51)(2)   & -3132.41488\,(72)      &  -3093.27246\,(47)     &  61.90632\,(54)(2)     &  101.04873\,(49)(2) \\
 33 &  75  & As   & 4.097\,(2)  & -3411.67124\,(54)(3)   & -3347.46900\,(69)      &  -3302.78726\,(48)     &  64.20224\,(54)(3)     &  108.88399\,(53)(3) \\
 34 &  80  & Se   & 4.140\,(2)  & -3636.55309\,(58)(3)   & -3570.02944\,(68)      &  -3519.22987\,(49)     &  66.52364\,(56)(3)     &  117.32322\,(56)(3) \\
 35 &  79  & Br   & 4.163\,(2)  & -3869.01443\,(60)(4)   & -3800.14450\,(67)      &  -3742.60759\,(49)     &  68.86993\,(57)(4)     &  126.40684\,(60)(4) \\
 36 &  84  & Kr   & 4.188\,(2)  & -4109.11165\,(64)(5)   & -4037.86739\,(67)      &  -3972.93073\,(50)     &  71.24426\,(60)(5)     &  136.18092\,(64)(5) \\
 37 &  85  & Rb   & 4.204\,(2)  & -4356.89633\,(68)(6)   & -4283.25052\,(65)      &  -4210.20713\,(51)     &  73.64581\,(62)(6)     &  146.68920\,(69)(6) \\
 38 &  88  & Sr   & 4.224\,(2)  & -4612.42740\,(73)(6)   & -4536.35074\,(66)      &  -4454.44706\,(52)     &  76.07665\,(66)(6)     &  157.98034\,(74)(6) \\
 39 &  89  & Y    & 4.243\,(2)  & -4875.76288\,(78)(7)   & -4797.22572\,(69)      &  -4705.65974\,(53)     &  78.53716\,(71)(7)     &  170.10314\,(79)(7) \\
 40 &  90  & Zr   & 4.269\,(1)  & -5146.96480\,(83)(7)   & -5065.93620\,(77)      &  -4963.85547\,(54)     &  81.02860\,(78)(7)     &  183.10933\,(84)(7) \\
 41 &  93  & Nb   & 4.324\,(2)  & -5426.09686\,(92)(9)   & -5342.5452\,(10)       &  -5229.04482\,(55)     &  83.5517\,(10)(1)      &  197.05205\,(94)(9) \\
 42 &  98  & Mo   & 4.409\,(2)  & -5713.2253\,(10)(1)    & -5627.1181\,(12)       &  -5501.23854\,(57)     &  86.1072\,(13)(1)      &  211.9868\,(10)(1) \\
 43 &  98  & Tc   & 4.42\,(4)   & -6008.4204\,(11)(19)   & -5919.7219\,(14)       &  -5780.44665\,(59)(4)  &  88.6985\,(15)(19)     &  227.9737\,(11)(19) \\
 44 & 102  & Ru   & 4.481\,(2)  & -6311.7525\,(12)(2)    & -6220.4280\,(15)       &  -6066.68168\,(61)     &  91.3244\,(16)(2)      &  245.0708\,(12)(2) \\
 45 & 103  & Rh   & 4.495\,(2)  & -6623.2973\,(13)(2)    & -6529.3087\,(16)       &  -6359.95425\,(63)     &  93.9886\,(17)(2)      &  263.3431\,(13)(2) \\
 46 & 106  & Pd   & 4.532\,(3)  & -6943.1300\,(14)(2)    & -6846.4400\,(15)       &  -6660.27692\,(66)     &  96.6900\,(16)(2)      &  282.8530\,(14)(2) \\
 47 & 107  & Ag   & 4.545\,(3)  & -7271.3321\,(15)(3)    & -7171.9003\,(14)       &  -6967.66147\,(69)(1)  &  99.4318\,(16)(3)      &  303.6707\,(15)(3) \\
 48 & 112  & Cd   & 4.594\,(2)  & -7607.9847\,(16)(3)    & -7505.7722\,(13)       &  -7282.12136\,(72)(1)  &  102.2125\,(16)(3)     &  325.8633\,(16)(3) \\
 49 & 115  & In   & 4.616\,(3)  & -7953.1761\,(18)(4)    & -7848.1391\,(13)       &  -7603.66861\,(75)(1)  &  105.0370\,(17)(4)     &  349.5075\,(18)(4) \\
 50 & 120  & Sn   & 4.652\,(2)  & -8306.9932\,(19)(4)    & -8199.0894\,(13)       &  -7932.31702\,(78)(1)  &  107.9037\,(19)(4)     &  374.6762\,(19)(4) \\
 51 & 121  & Sb   & 4.680\,(3)  & -8669.5279\,(21)(5)    & -8558.7129\,(14)       &  -8268.07926\,(81)(1)  &  110.8150\,(21)(5)     &  401.4486\,(21)(5) \\
 52 & 130  & Te   & 4.742\,(3)  & -9040.8755\,(23)(5)    & -8927.1058\,(15)       &  -8610.97093\,(85)(1)  &  113.7697\,(23)(5)     &  429.9046\,(23)(5) \\
 53 & 127  & I    & 4.750\,(8)  & -9421.1396\,(25)(13)   & -9304.3633\,(16)       &  -8961.00370\,(89)(2)  &  116.7763\,(25)(13)    &  460.1359\,(25)(13) \\
 54 & 132  & Xe   & 4.786\,(5)  & -9810.4189\,(27)(10)   & -9690.5894\,(17)       &  -9318.19469\,(93)(1)  &  119.8295\,(27)(10)    &  492.2242\,(27)(10) \\
 55 & 133  & Cs   & 4.804\,(5)  & -10208.8228\,(30)(11)  & -10085.8875\,(18)      &  -9682.55729\,(98)(2)  &  122.9352\,(30)(11)    &  526.2655\,(30)(11) \\
 56 & 138  & Ba   & 4.838\,(5)  & -10616.4583\,(33)(12)  & -10490.3674\,(19)      &  -10054.1077\,(10)     &  126.0909\,(33)(12)    &  562.3505\,(33)(12) \\
 57 & 139  & La   & 4.855\,(5)  & -11033.4441\,(36)(14)  & -10904.1415\,(21)      &  -10432.8607\,(11)     &  129.3026\,(36)(14)    &  600.5834\,(36)(14) \\
 58 & 140  & Ce   & 4.877\,(2)  & -11459.8962\,(39)(11)  & -11327.3277\,(22)      &  -10818.8327\,(11)     &  132.5685\,(39)(11)    &  641.0635\,(39)(11) \\
 59 & 141  & Pr   & 4.892\,(5)  & -11895.9407\,(43)(18)  & -11760.0475\,(23)      &  -11212.0401\,(12)     &  135.8932\,(43)(18)    &  683.9006\,(43)(18) \\
 60 & 142  & Nd   & 4.912\,(3)  & -12341.7017\,(46)(15)  & -12202.4271\,(24)      &  -11612.4997\,(13)     &  139.2746\,(46)(15)    &  729.2020\,(46)(15) \\
 61 & 145  & Pm   & 4.96\,(5)   & -12797.3042\,(52)(176) & -12654.5968\,(26)(4)   &  -12020.2286\,(13)(2)  &  142.7074\,(52)(176)   &  777.0756\,(51)(176) \\
 62 & 152  & Sm   & 5.082\,(6)  & -13262.8663\,(57)(29)  & -13116.6931\,(28)(1)   &  -12435.2451\,(14)     &  146.1732\,(57)(29)    &  827.6212\,(56)(29) \\
 63 & 153  & Eu   & 5.112\,(6)  & -13738.5877\,(62)(34)  & -13588.8569\,(30)(1)   &  -12857.5649\,(15)     &  149.7308\,(63)(34)    &  881.0227\,(62)(34) \\
 64 & 158  & Gd   & 5.157\,(4)  & -14224.5802\,(66)(31)  & -14071.2351\,(31)(1)   &  -13287.2080\,(16)     &  153.3451\,(67)(31)    &  937.3722\,(66)(31) \\
 65 & 159  & Tb   & 5.1\,(2)    & -14721.0774\,(73)(823) & -14563.9810\,(33)(24)  &  -13724.1910\,(17)(10) &  157.0965\,(74)(823)   &  996.8865\,(73)(823) \\
 66 & 162  & Dy   & 5.21\,(2)   & -15228.0330\,(79)(110) & -15067.2462\,(36)(3)   &  -14168.5358\,(17)(1)  &  160.7868\,(80)(110)   &  1059.4972\,(79)(110) \\
 67 & 165  & Ho   & 5.20\,(3)   & -15745.8310\,(88)(215) & -15581.2024\,(38)(7)   &  -14620.2592\,(18)(2)  &  164.6286\,(90)(215)   &  1125.5718\,(88)(215) \\
 68 & 166  & Er   & 5.252\,(3)  & -16274.5200\,(96)(44)  & -16106.0158\,(41)(1)   &  -15079.3822\,(19)     &  168.5042\,(98)(44)    &  1195.1378\,(96)(44) \\
 69 & 169  & Tm   & 5.226\,(4)  & -16814.375\,(10)(5)    & -16641.8677\,(43)(2)   &  -15545.9243\,(21)(1)  &  172.507\,(11)(5)      &  1268.450\,(10)(5) \\
 70 & 174  & Yb   & 5.311\,(6)  & -17365.429\,(11)(7)    & -17188.9366\,(46)(3)   &  -16019.9078\,(22)(1)  &  176.492\,(11)(7)      &  1345.521\,(11)(7) \\
 71 & 175  & Lu   & 5.37\,(3)   & -17927.979\,(12)(32)   & -17747.4159\,(49)(13)  &  -16501.3515\,(23)(3)  &  180.563\,(13)(32)     &  1426.628\,(13)(32) \\
 72 & 180  & Hf   & 5.347\,(3)  & -18502.302\,(14)(7)    & -18317.5115\,(53)(3)   &  -16990.2773\,(24)(1)  &  184.790\,(14)(7)      &  1512.024\,(14)(7) \\
 73 & 181  & Ta   & 5.351\,(3)  & -19088.493\,(15)(8)    & -18899.4254\,(56)(4)   &  -17486.7072\,(26)(1)  &  189.068\,(15)(8)      &  1601.786\,(15)(8) \\
 74 & 184  & W    & 5.366\,(2)  & -19686.785\,(16)(9)    & -19493.3753\,(60)(4)   &  -17990.6640\,(27)(1)  &  193.410\,(17)(9)      &  1696.121\,(16)(9) \\
 75 & 187  & Re   & 5.37\,(2)   & -20297.433\,(18)(28)   & -20099.5877\,(64)(14)  &  -18502.1698\,(29)(3)  &  197.845\,(18)(28)     &  1795.263\,(18)(28) \\
 76 & 192  & Os   & 5.413\,(2)  & -20920.591\,(19)(10)   & -20718.2926\,(69)(5)   &  -19021.2485\,(30)(1)  &  202.299\,(20)(11)     &  1899.343\,(19)(10) \\
 77 & 193  & Ir   & 5.4\,(1)    & -21556.655\,(21)(204)  & -21349.7439\,(73)(112) &  -19547.9219\,(32)(18) &  206.911\,(21)(205)    &  2008.733\,(21)(204) \\
 78 & 196  & Pt   & 5.431\,(3)  & -22205.723\,(22)(14)   & -21994.1897\,(78)(8)   &  -20082.2159\,(34)(1)  &  211.533\,(22)(14)     &  2123.507\,(22)(14) \\
 79 & 197  & Au   & 5.437\,(4)  & -22868.180\,(25)(17)   & -22651.9033\,(84)(10)  &  -20624.1537\,(36)(1)  &  216.277\,(25)(17)     &  2244.027\,(25)(17) \\
 80 & 202  & Hg   & 5.465\,(3)  & -23544.197\,(25)(19)   & -23323.1577\,(90)(11)  &  -21173.7616\,(38)(2)  &  221.039\,(26)(19)     &  2370.436\,(25)(19) \\
 81 & 205  & Tl   & 5.476\,(3)  & -24234.170\,(29)(20)   & -24008.2503\,(97)(13)  &  -21731.0636\,(40)(2)  &  225.920\,(30)(20)     &  2503.106\,(29)(20) \\
 82 & 208  & Pb   & 5.501\,(1)  & -24938.297\,(30)(22)   & -24707.480\,(10)(1)    &  -22296.0864\,(42)(2)  &  230.817\,(31)(22)     &  2642.210\,(30)(22) \\
 83 & 209  & Bi   & 5.521\,(3)  & -25656.972\,(35)(26)   & -25421.170\,(11)(2)    &  -22868.8557\,(45)(2)  &  235.802\,(36)(26)     &  2788.116\,(35)(26) \\
 84 & 209  & Po   & 5.53\,(2)   & -26390.543\,(38)(74)   & -26149.657\,(12)(5)    &  -23449.3982\,(47)(6)  &  240.886\,(39)(74)     &  2941.145\,(38)(74) \\
 85 & 210  & At   & 5.54\,(6)   & -27139.291\,(41)(241)  & -26893.287\,(13)(18)   &  -24037.7420\,(50)(18) &  246.004\,(42)(242)    &  3101.549\,(41)(241) \\
 86 & 222  & Rn   & 5.69\,(2)   & -27902.937\,(45)(104)  & -27652.375\,(14)(8)    &  -24633.9209\,(53)(8)  &  250.562\,(47)(104)    &  3269.016\,(45)(104) \\
 87 & 223  & Fr   & 5.70\,(2)   & -28683.153\,(49)(103)  & -28427.408\,(16)(8)    &  -25237.9510\,(56)(7)  &  255.746\,(51)(103)    &  3445.202\,(49)(103) \\
 88 & 226  & Ra   & 5.72\,(3)   & -29479.573\,(53)(176)  & -29218.731\,(17)(15)   &  -25849.8684\,(60)(12) &  260.842\,(55)(176)    &  3629.704\,(54)(176) \\
 89 & 227  & Ac   & 5.67\,(6)   & -30293.205\,(58)(373)  & -30026.820\,(19)(33)   &  -26469.6981\,(63)(25) &  266.385\,(60)(374)    &  3823.507\,(58)(373) \\
 90 & 232  & Th   & 5.78\,(1)   & -31122.902\,(57)(105)  & -30851.981\,(20)(10)   &  -27097.4807\,(66)(7)  &  270.921\,(60)(106)    &  4025.421\,(58)(105) \\
 91 & 231  & Pa   & 5.70\,(6)   & -31971.651\,(69)(459)  & -31694.914\,(22)(44)   &  -27733.2297\,(70)(30) &  276.737\,(72)(462)    &  4238.421\,(69)(459) \\
 92 & 238  & U    & 5.857\,(3)  & -32836.564\,(69)(34)   & -32555.797\,(23)(4)    &  -28376.9990\,(74)     &  280.767\,(72)(34)     &  4459.565\,(69)(34) \\
 93 & 237  & Np   & 5.74\,(6)   & -33722.354\,(82)(568)  & -33435.562\,(26)(58)   &  -29028.7916\,(78)(35) &  286.792\,(86)(571)    &  4693.562\,(82)(568) \\
 94 & 244  & Pu   & 5.89\,(4)   & -34624.711\,(81)(427)  & -34334.273\,(28)(45)   &  -29688.6715\,(83)(26) &  290.438\,(85)(430)    &  4936.040\,(81)(427) \\
 95 & 243  & Am   & 5.905\,(4)  & -35548.087\,(99)(108)  & -35252.912\,(31)(11)   &  -30356.6463\,(88)(6)  &  295.17\,(10)(11)      &  5191.440\,(99)(108) \\
 96 & 246  & Cm   & 5.86\,(2)   & -36492.456\,(95)(270)  & -36192.041\,(33)(31)   &  -31032.7518\,(96)(16) &  300.42\,(10)(27)      &  5459.705\,(95)(270) \\
 97 & 247  & Bk   & 5.82\,(6)   & -37457.73\,(12)(87)    & -37152.237\,(38)(103)  &  -31717.030\,(11)(5)   &  305.49\,(12)(87)      &  5740.70\,(12)(87) \\
 98 & 252  & Cf   & 5.85\,(6)   & -38443.41\,(11)(97)    & -38134.049\,(41)(119)  &  -32409.521\,(11)(5)   &  309.36\,(12)(98)      &  6033.88\,(11)(97) \\
 99 & 252  & Es   & 5.85\,(6)   & -39451.79\,(14)(107)   & -39138.401\,(46)(136)  &  -33110.251\,(11)(6)   &  313.39\,(15)(108)     &  6341.54\,(14)(107) \\
100 & 253  & Fm   & 5.86\,(6)   & -40482.93\,(16)(119)   & -40165.992\,(51)(157)  &  -33819.260\,(12)(6)   &  316.94\,(16)(120)     &  6663.67\,(16)(119) \\
 \hline\hline\\[-5pt]
\end{longtable*}

%
%
%
\begin{table*}[htb]
\begin{center}
\caption{
Comparison of
the approximate treatment of QED effects based on the model QED operator ($E_{\rm mqed}$)
with {\em ab initio} QED calculations. $E_{\rm qed,rad}$ denotes
the radiative QED correction (including the two-loop effects), whereas
$E^{(1)}_{\rm qed}$ and $E^{(2)}_{\rm qed}$ are the one-photon and two-photon electron-structure  QED
corrections, respectively.
Units are eV.
\label{tab:qedmod}
}
\begin{ruledtabular}
\begin{tabular}{lw{2.4}w{2.6}w{1.1}w{2.4}w{1.1}w{2.4}w{1.1}}
$Z$
   & \multicolumn{1}{c}{$E_{\rm mqed}$} & \multicolumn{1}{c}{$E_{\rm qed,rad}$}
                               & \multicolumn{1}{c}{$\frac{E_{\rm qed,rad}-E_{\rm mqed}}{E_{\rm mqed}} [\%]$}
                                          & \multicolumn{1}{c}{$E^{(1)}_{\rm qed}$}
                               & \multicolumn{1}{c}{$\frac{E^{(1)}_{\rm qed}}{E_{\rm mqed}} [\%]$}
                                           & \multicolumn{1}{c}{$E^{(2)}_{\rm qed}$}
                               & \multicolumn{1}{c}{$\frac{E^{(2)}_{\rm qed}}{E_{\rm mqed}} [\%]$}
\\\hline\\[-9pt]
%
%
%
  \multicolumn{5}{l}{$2p_{1/2}$--$2s$} \\
 10   &    -0.01404    &   -0.01417\,(6)   &    1.0     &     -0.00001  &    0.1      &    -0.00007  &    0.5 \\
 20   &    -0.1911     &   -0.1920\,(2)    &    0.5     &     -0.0006   &    0.3      &    -0.0005   &    0.3 \\
 30   &    -0.8122     &   -0.8143\,(3)    &    0.3     &     -0.0045   &    0.5      &    -0.0015   &    0.2 \\
 40   &    -2.2207     &   -2.2227\,(6)    &    0.1     &     -0.0187   &    0.8      &    -0.0033   &    0.1 \\
 50   &    -4.820      &   -4.817\,(2)     &   -0.1     &     -0.054    &    1.1      &    -0.005    &    0.1 \\
 60   &    -9.092      &   -9.076\,(4)     &   -0.2     &     -0.122    &    1.3      &    -0.008    &    0.1 \\
 70   &   -15.62       &   -15.58\,(1)     &   -0.3     &     -0.23     &    1.4      &    -0.01     &    0.1 \\
 83   &   -28.69       &   -28.59\,(3)     &   -0.3     &     -0.36     &    1.3      &    -0.01     &    0.0 \\
 92   &   -41.65       &   -41.51\,(7)     &   -0.3     &     -0.34     &    0.8      &    -0.01     &    0.0 \\
%
%
  \multicolumn{5}{l}{$2p_{3/2}$--$2s$} \\
 10   &    -0.01336    &   -0.01368\,(6)   &    2.3     &     -0.00006  &    0.4      &    -0.00008   &   0.6 \\
 20   &    -0.1778     &   -0.1804\,(2)    &    1.5     &     -0.0027   &    1.5      &    -0.0008    &   0.4 \\
 30   &    -0.7429     &   -0.7504\,(3)    &    1.0     &     -0.0228   &    3.1      &    -0.0030    &   0.4 \\
 40   &    -2.0050     &   -2.0184\,(7)    &    0.7     &     -0.1011   &    5.0      &    -0.0083    &   0.4 \\
 50   &    -4.315      &   -4.332\,(2)     &    0.4     &     -0.317    &    7.3      &    -0.018     &   0.4 \\
 60   &    -8.121      &   -8.129\,(4)     &    0.1     &     -0.801    &    9.9      &    -0.037     &   0.4 \\
 70   &   -14.03       &   -14.01\,(1)     &   -0.2     &     -1.75     &   12.5      &    -0.07      &   0.5 \\
 83   &   -26.37       &   -26.23\,(3)     &   -0.5     &     -4.12     &   15.6      &    -0.13      &   0.5 \\
 92   &   -39.46       &   -39.15\,(7)     &   -0.8     &     -6.88     &   17.4      &    -0.20      &   0.5 \\
%
%
  \multicolumn{5}{l}{$2p_{3/2}$--$2p_{1/2}$} \\
 10   &     0.00068    &    0.00050\,(3)   &   -26.3     &     -0.00005  &   -7       &    -0.00001   &  -2.1 \\
 20   &     0.0132     &    0.0116\,(1)    &   -12.5     &     -0.0021   &   -16      &    -0.0003    &  -1.9 \\
 30   &     0.0692     &    0.0638\,(2)    &    -7.7     &     -0.0183   &   -26      &    -0.0015    &  -2.1 \\
 40   &     0.2157     &    0.2042\,(4)    &    -5.3     &     -0.0823   &   -38      &    -0.0050    &  -2.3 \\
 50   &     0.5045     &    0.4856\,(5)    &    -3.7     &     -0.2627   &   -52      &    -0.0132    &  -2.6 \\
 60   &     0.971      &    0.947\,(1)     &    -2.5     &     -0.679    &   -70      &    -0.029     &  -3.0 \\
 70   &     1.593      &    1.574\,(3)     &    -1.2     &     -1.523    &   -95      &    -0.057     &  -3.6 \\
 83   &     2.322      &    2.364\,(9)     &     1.8     &     -3.755    &   -162     &    -0.121     &  -5.2 \\
 92   &     2.18       &    2.36\,(2)      &     7.9     &     -6.53     &   -299     &    -0.19      &  -8.9 \\
\end{tabular}
\end{ruledtabular}
\end{center}
\end{table*}


\end{document}